\documentclass[12pt]{iopart}
\usepackage{iopams}
\usepackage{epsfig,bm,epsf,float}
\usepackage{latexsym}

\begin{document}

\newcommand{\tbox}[1]{\mbox{\tiny #1}}
\newcommand{\halfroot}{\mbox{\small $\frac{1}{\sqrt{2}}$}}
\newcommand{\eexp}[1]{{e}^{#1}}
\newcommand{\half}{\mbox{\small $\frac{1}{2}$}}
\newcommand{\pit}{\mbox{\small $\frac{\pi}{2}$}}
\newcommand{\mbf}[1]{{\mathbf #1}}
\def\abs#1{\left| #1 \right|}
\newcommand{\op}[1]{\hat{\mbf{#1}}}

\newcommand{\nssec}{\,\mbox{ns}}
\newcommand{\mcsec}{\,\mu\mbox{s}}
\newcommand{\msec}{\,\mbox{ms}}
\newcommand{\seconds}{\,\mbox{s}}
\newcommand{\hz}{\,\mbox{Hz}}
\newcommand{\khz}{\,\mbox{KHz}}
\newcommand{\Mhz}{\,\mbox{MHz}}
\newcommand{\ghz}{\,\mbox{GHz}}
\newcommand{\nm}{\,\mbox{nm}}
\newcommand{\mcm}{\,\mu\mbox{m}}
\newcommand{\mm}{\,\mbox{mm}}
\newcommand{\cm}{\,\mbox{cm}}
\newcommand{\m}{\,\mbox{m}}
\newcommand{\nw}{\,\mbox{nW}}
\newcommand{\mcw}{\,\mu\mbox{W}}
\newcommand{\mw}{\,\mbox{mW}}
\newcommand{\w}{\,\mbox{W}}
\newcommand{\mck}{\,\mu\mbox{K}}
\newcommand{\mk}{\,\mbox{mK}}
\newcommand{\mg}{\,\mbox{mG}}
\newcommand{\g}{\,\mbox{G}}
\newcommand{\vrc}{\,v_\textrm{rec}}
\newcommand{\erc}{\,E_\textrm{rec}}
\newcommand{\ohf}{\omega_{\tbox{HF}}}
\newcommand{\of}{\omega_{\tbox{F}}}
\newcommand{\ehf}{E_{\tbox{HF}}}
\newcommand{\omw}{\omega_{\tbox{MW}}}
\newcommand{\Omw}{\Omega_{\tbox{MW}}}
\newcommand{\vmw}{V_{\tbox{MW}}}
\newcommand{\Dmw}{\Delta_{\tbox{MW}}}

\newcommand{\bra}[1]{\left\langle #1\right|}
\newcommand{\ket}[1]{\left|#1\right\rangle}
\newcommand{\braket}[2]{\left\langle #1 \mid #2 \right\rangle}
\newcommand{\braOket}[3]{\left\langle #1\left|#2\right|#3\right\rangle}
\newcommand{\da}{\downarrow}
\newcommand{\ua}{\uparrow}

\title{Hyperfine Spectroscopy of Optically Trapped Atoms}

\author{A. Kaplan, M. F. Andersen, T. Gr\"{u}nzweig, and N. Davidson\footnote{To
whom correspondence should be addressed (fedavid@wisemail.weizmann.ac.il).} }

\address{Department of Physics of Complex Systems,
Weizmann Institute of Science, Rehovot, Israel.}

\begin{abstract}
We perform spectroscopy on the hyperfine splitting of $^{85}$Rb atoms trapped in far-off-resonance optical traps.
The existence of a spatially dependent shift in the energy levels is shown to induce an inherent dephasing effect,
which causes a broadening of the spectroscopic line and hence an inhomogeneous loss of atomic coherence at a much
faster rate than the homogeneous one caused by spontaneous photon scattering. We present here a number of
approaches for reducing this inhomogeneous broadening, based on trap geometry, additional laser fields, and novel
microwave pulse sequences. We then show how hyperfine spectroscopy can be used to study quantum dynamics of
optically trapped atoms.

\end{abstract}




\section{Introduction} \label{ch:intro}

The development of laser cooling and trapping techniques \cite{Cohen-Tannoudji98, Phillips98, Chu98} has been
motivated, to a great extent, by their potential use for precision measurements, of which precision frequency
measurements are of outmost importance. Cold atoms are attractive for these measurements mainly because of the
suppression of the Doppler broadening and because they allow for a longer interrogation time and hence a narrower
linewidth.

Modern atomic frequency standards are based on atomic clocks, which measure the frequency associated with the
transition between two ground state hyperfine levels. The performance of such a clock can be improved by
lengthening the time over which such a transition is measured. During the 1950's various possibilities were
suggested to lengthen this interrogation time: Ramsey \cite{Ramsey90} proposed confining the atoms in a storage
ring with an inhomogeneous magnetic field, or a storage cell with properly coated walls. Zacharias
\cite{Zacharias} proposed to build an atomic ``fountain'' in which a thermal beam of atoms is launched upward, so
slow atoms from the tail of the velocity distribution will turn around and fall back under the force of gravity.
Using the Ramsey method of separated fields, the atoms could be interrogated once on their way up and then again
on their way down, resulting in a long total interrogation time. Back then, this idea failed because of collisions
of these slow atoms with the faster ones. Using laser-cooled atoms, Zacharias' idea was revived in 1989 by
Kasevich et. al. \cite{Kasevich89}. They cooled sodium atoms using a magneto-optical trap \cite{Raab87} and
launched them on ballistic trajectories by applying a short pulse of resonant light. Two short $\pit$ microwave
pulses were applied as the atoms turned around inside a waveguide and allowed the ground-state hyperfine splitting
to be measured with a linewidth of $2 \hz$.

To further increase the measurement time beyond the practical limit imposed by the height of a fountain, one can
use trapped atoms, and in particular optical dipole traps, based on the dipole potential created by the
interaction of a neutral atom and a laser beam \cite{Chu86,Grimm00}. Unfortunately, the interaction between the
trapping light and the trapped atoms is not negligible. First, even for a far-off-resonance trap, where many of
the undesired effects of the atom-light interaction are suppressed, residual spontaneous scattering of photons
from the trap laser destroys the ground state spin polarization or spin coherence \cite{Cline94}. Next, and more
importantly, atom trapping relies on inducing a spatial inhomogeneous shift of the atomic energy levels, and the
existence of such a  shift makes precision spectroscopy of trapped atoms a difficult task. In the case of
hyperfine spectroscopy in optical traps, the inhomogeneity is approximately equal for both states, since the
trapping light interacts with both levels with the same strength (the dipole matrix elements are identical).
However, the difference in detuning between them results in a \emph{differential} inhomogeneity, which induces a
dephasing effect and, when ensemble-averaged, causes a broadening of the spectroscopic line and hence a loss of
atomic coherence at a much faster rate than the spontaneous photon scattering rate \cite{Davidson95}. These
effects can be reduced by increasing the trap detuning \cite{Davidson95,Adams95,Miller93,Takekoshi96} and also by
using blue-detuned optical traps, in which the atoms are confined mainly in the dark \cite{Davidson95,Friedman02}.
However, the residual frequency shifts are still the main factor limiting the coherence of the trapped atoms, and
hence the use of dipole traps for precision spectroscopy.

A number of approaches for reducing the inhomogeneous broadening of the ground state hyperfine splitting of
optically trapped atoms are presented in this tutorial. Following a brief review on light shifts and optical
dipole traps in section \ref{ch:back}, and a description of the basic experimental setup in section
\ref{ap:setup}, we present in section \ref{sec:classical} a simple, classical model for microwave spectroscopy in
an optical trap, and use it to analyze the atomic coherence times for different trap geometries in section
\ref{sec:case}. Based on this model, we present in section \ref{sec:compensating} a novel method for reducing the
inhomogeneous frequency broadening by the addition of a weak light field, spatially mode-matched with the trapping
field and whose frequency is tuned in-between the two hyperfine levels. In section \ref{sec:selection} we show
another way of achieving long coherence times in an optical trap: a narrow energy distribution is achieved using a
microwave ``pre-selection'' pulse which selects a subset of the atomic ensemble. Our technique allows the
selection of a narrow energy band around any central energy enabling us to maximize the number of selected atoms
(for a given energy width) by choosing the energy with the highest density of populated states.

A full, quantum-mechanical model is developed in section \ref{sec:quantum}. The limit of short and strong pulses
is further elaborated in section \ref{sec:ramsey}, and used to predict the Ramsey decoherence time of a thermal
ensemble as a consequence of dephasing. In section \ref{sec:echo} it is shown that, for certain trap parameters,
the dephasing can be reversed by stimulating a ``coherence echo''. The failure of the echo for other trap
parameters is due to dynamics in the trap, and thereby ``echo spectroscopy'' can also be used to study quantum
dynamics in the trap even when more than 10$^{6}$ states are thermally populated, and to study the crossover from
quantum dynamics to classical dynamics. We then show (in section \ref{ssec:manypi}) that the decay in the
hyperfine coherence due to interactions with the environment, is only partly suppressed by echo spectroscopy,
primarily due to dynamical (time-dependent) dephasing mechanisms. An improved pulse sequence is demonstrated,
containing additional $\pi$ pulses for which the decay of coherence is reduced by a factor 2.5 beyond the
reduction offered by the simple echo scheme.

Finally, in section \ref{sec:chaos} we show that our echo spectroscopy methods enable us to study quantum dynamics
of trapped atoms also for chaotic and mixed dynamics where, surprisingly, partial revivals of atomic coherence
occur in the perturbative (quantum) regime, and disappear in the non-perturbative (semi-classical) regime,
indicating a clear quantum to classical transition as a function of the perturbation strength.

\section{Light Shifts and Dipole Traps} \label{ch:back}
Throughout the work presented in this tutorial we use far-off-resonance optical dipole traps, which are traps
based on the dipole potential created by the interaction of a neutral atom and a laser beam \cite{Grimm00}. A few
concepts on which the rest of the work presented here is based, are reviewed. In sections \ref{sec:potential} and
\ref{sec:shift} a brief review of the origin of the dipole potential, and its connection with the ac Stark shift
of the ground-state is presented using an idealized two-level atom. In section \ref{sec:trapping} a short review
on the use of the dipole potential to trap neutral atoms is given\footnote{Laser-cooling of the atoms is a
prerequisite for dipole trapping, and is done with techniques that have became standard in many research
laboratories, and will not be reviewed here (See Refs. \cite{Cohen-Tannoudji90,Phillips90,Adams97}).}. In sections
\ref{ssec:back-scatter} and \ref{ssec:shift-multi}, a description of the main mechanisms for atomic decoherence is
given, this time for multi-level atoms.

\subsection{The Optical Dipole Potential} \label{sec:potential}

The interaction of a neutral atom and a nearly-resonant electro-magnetic field is governed by the dipole
interaction, and is usually separated into two terms which correspond to a \emph{reactive} force and a
\emph{dissipative} force. When an atom is exposed to light, the electric field component induces a dipole moment
in the atom, oscillating at the driving-light frequency.

If we assume for example a monochromatic light field with an electric field given by $\vec{E}(\vec{r},t) = \hat{e}
E(\vec{r}) \eexp{-i \omega_L t} + c.c.$, where $\hat{e}$ is the unit polarization vector and $\omega_L$ the
frequency, then the dipole moment is $\vec{d}(\vec{r},t)=\hat{e} d(\vec{r}) \eexp{-i \omega_L t} + c.c. $. The
amplitude of the dipole moment is related to the amplitude of the field by $d= \alpha E$, where $\alpha$ is the
atomic complex polarizability, which is a function of the driving frequency. The interaction of the induced dipole
moment with the driving field gives rise to the potential \cite{Grimm00}
\begin{eqnarray}
U_{dip}=-\frac{1}{2}\left\langle\vec{d}\vec{E}\right\rangle=-\frac{1}{2 \epsilon_0 c} \mathrm{Re}(\alpha)
I(\vec{r}),
\end{eqnarray}
where $I=2 \epsilon_0 c \left|E \right|^2$ is the light intensity. The reactive ``dipole'' force is a conservative
one, and is equal to the gradient of the above potential. The dissipative force is related to the power the
oscillating dipole absorbs from the field, which is given by:
\begin{eqnarray}
P_{diss}=\left\langle\dot{\vec{d}}\vec{E}\right\rangle=\frac{\omega_L}{\epsilon_0 c} \mathrm{Im}(\alpha)
I(\vec{r}). \label{diss}\end{eqnarray}

In a quantum picture, the dipole force results from absorption of a photon from the field followed by stimulated
emission of a photon into a different mode of the laser field. The momentum transfer is the vector difference
between the momenta of the absorbed and emitted photons. The dissipative component has its origin in cycles of
absorption of photons, followed by spontaneous emission, in a random direction. Using equation \ref{diss} we can
write an equation for the rate of spontaneous photon scattering:
\begin{eqnarray}
\gamma_s=\frac{P_{diss}}{\hbar \omega_L}=\frac{1}{\hbar \epsilon_0 c} \mathrm{Im}(\alpha) I(\vec{r}).
\end{eqnarray}

The atomic polarizability can be calculated by using the solutions of the optical Bloch equations, while
translational degrees of freedom are taken into account \cite{Cohen-Tannoudji92}. For a two-level atom with
resonance frequency $\omega_0$, and using the ``rotating wave approximation'', the average dipole potential can be
written as
\begin{eqnarray}
U_{dip}\left( \vec{r}\right) =\frac{\hbar \delta }{2 } \mathrm{ln}\left(1+ \frac{I/I_0}{1 + 4 (\delta / \gamma)^2}
\right) \label{ldip}.
\end{eqnarray}
The resulting expression for the average scattering rate is
\begin{eqnarray}
\gamma_s=\frac{\gamma }{2}\left( \frac{I/I_0}{1+I/I_0+4 (\delta / \gamma)^2}\right) \label{lscat},
\end{eqnarray}
where $\gamma $ is the natural linewidth of the atomic transition, $\delta =\omega_{L}-\omega_0$ is the detuning
of the laser from the atomic resonance, $I_0$ is the saturation intensity of the transition, given by $I_{0}=2\pi
^{2}\hbar \gamma c/3\lambda _{0}^{3}$, and $\lambda _{0}=2 \pi c /\omega _{0}$. For a large detuning from
resonance ($\delta \gg \gamma$) equations \ref{ldip} and \ref{lscat} can be approximated as
\begin{eqnarray}
U_{dip}\left( \vec{r}\right) &=&\frac{3\pi c^{2}}{2\omega _{0}^{3}}\frac{\gamma }{\delta }I\left( \vec{r}\right),\label{sdip}\\
\gamma_s\left( \vec{r}\right) &=&\frac{3\pi c^{2}}{2 \hbar \omega _{0}^{3}}\left(\frac{\gamma }{\delta }\right)^2
I\left( \vec{r}\right).\label{sscat}
\end{eqnarray}

\subsection{Ground State Light Shifts} \label{sec:shift}

\begin{figure}[tb]
\begin{center}
\includegraphics[width=5in] {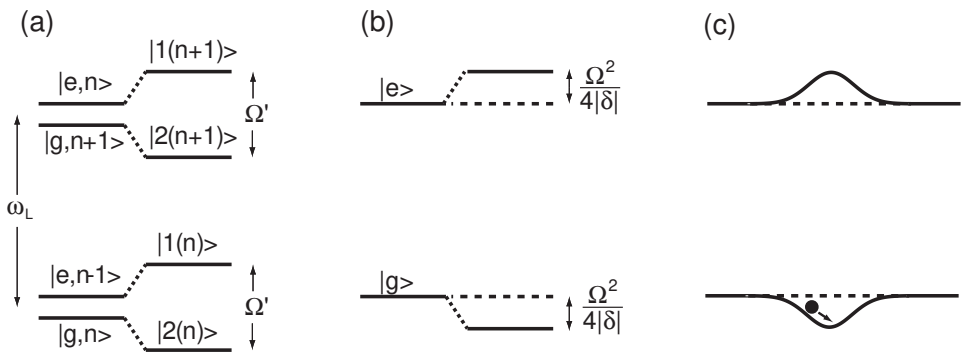}
\caption{Dressed state picture of a two-level atom coupled by a light field with frequency below the atomic
resonance. (a) The energy levels form a ladder of (almost degenerate) manifolds separated by the photon energy
$\hbar \omega_L$. The atom-field interaction separates the levels within each manifold by $\hbar \Omega'$, where
$\Omega'$ is the effective Rabi frequency. The eigenstates of the coupled system are a mixture of the eigenstates
of the uncoupled system. (b) If $\Omega \ll \abs{\delta}$, then the dressed eigenstates can be identified with the
original ground and excited states, with a small portion of the other state. The energy shift is $\hbar \Omega^2/4
\delta $. (c) A spatially inhomogeneous light field produces a ground-state potential well, in which the atom
(shown schematically for a Gaussian trap) can be trapped.} \label{dressed}
\end{center}
\end{figure}

It is useful to look at the atom-light interaction in the ``dressed state'' model \cite{Cohen-Tannoudji92}. The
combined Hamiltonian for the atom and the laser field is given by
$\mathcal{H}=\mathcal{H}_A+\mathcal{H}_L+\mathcal{H}_{AL}$, where $\mathcal{H}_A$,$\mathcal{H}_L$, and
$\mathcal{H}_{AL}$ are the atom, laser and interaction parts of the Hamiltonian, respectively. Consider a
two-level atom with ground state $\ket{g}$ and excited state $\ket{e}$ with energy $\hbar \omega_0$, in the
presence of a laser with frequency $\omega_L$. The atom can be described by the Hamiltonian
\begin{eqnarray}
\mathcal{H}_A=\hbar \omega_0 \ket{e} \bra{e},
\end{eqnarray}
and the laser field by
\begin{eqnarray}
\mathcal{H}_L=\hbar \omega_L \left( a^\dag a + \frac{1}{2} \right),
\end{eqnarray}
where $a^\dag $ and $a$ are the photon creation and annihilation operators, respectively. If the coupling between
the atom and the field is ignored, then the eigenstates of $\mathcal{H}_A+\mathcal{H}_L$ are characterized by the
atomic internal state (either $\ket{g}$ or $\ket{e}$) and by the number of photons in the electromagnetic field,
$n$. The energy levels form a ladder of manifolds, separated by $\hbar \omega_L$, each containing two states of
the form $\ket{g,n}$ and $\ket{e,n-1}$ (see figure \ref{dressed}a). If the light is in resonance with the atomic
transition, these two states are degenerate. Otherwise, they are separated by $\hbar \abs{\delta}$.

A general result from second-order time-independent perturbation theory is that an interaction Hamiltonian
$\mathcal{H'}$ will lead to a shift in the energy of the (unperturbed) $i$-th state given by \cite{Grimm00}
\begin{eqnarray}
\Delta E_i=\sum_{j\neq i} \frac{\abs{\braOket{j}{\mathcal{H'}}{i}}^2}{E_i-E_j}.
\end{eqnarray}
The interaction term in our case is $\mathcal{H}_{AL}=-\op{d} \cdot E$, where $\op{d}=-e \op{r}$ is the electric
dipole operator. This interaction term couples states within the same manifold with a matrix element given by
\begin{eqnarray}
\braOket{e,n-1}{H_{AL}}{g,n}=\frac{\hbar \Omega}{2},
\end{eqnarray}
where $\Omega=d E / \hbar$ is the Rabi frequency. The eigenfunctions of the coupled system, denoted ``dressed
states'', are a mixture of the uncoupled system eigenstates and are given by \cite{Adams97}
\begin{eqnarray}
\ket{1(n)}=\cos \theta \ket{e,n-1} - \sin \theta \ket{g,n} \\
\ket{2(n)}=\sin \theta \ket{e,n-1} + \cos \theta \ket{g,n},
\end{eqnarray}
where $\tan 2 \theta = -\Omega/\delta$. The new (dressed) levels are separated by an energy $\hbar \Omega'$, where
$\Omega'= \sqrt{\Omega^2+\delta^2}$ is called the ``generalized Rabi frequency''. For a large detuning, $\left|
\delta \right| \gg \Omega$, the ground state is shifted by $\hbar \Omega^2 / 4 \delta$ (see figure
\ref{dressed}b). The dressed states $\ket{2(n)}$,$\ket{2(n+1)}$, ... can be identified with the original ground
state, with a small mixture of the excited state. This light-induced shift in the atomic energy levels is usually
denoted ``light shift'' or ``ac Stark shift''. Using  $\Omega ^{2}=\left( \gamma ^{2}/2\right) \left( I/I_0\right)
$, where $I_0$ is the saturation intensity of the transition, we conclude that the light-shift of the ground state
energy, $\Delta E_g$, is given by
\begin{eqnarray}
\Delta E_g=\frac{\hbar \gamma^2}{8 \delta}\frac{I}{I_0}=\frac{3\pi c^{2}}{2\omega _{0}^{3}}\frac{\gamma }{\delta
}I\left( \vec{r}\right) \label{shift2}.
\end{eqnarray}
As seen, in the perturbative limit the light-shift of the ground state is equal to the dipole potential in
equation \ref{sdip}. The reason is that, since in this regime the atom is mostly in the ground state, the light
shifted ground state can be identified as the potential for the motion of the atoms (see figure \ref{dressed}c).
In the case of multi-level atoms, equation \ref{shift2} should be modified to include the electric dipole
interaction between the ground-state and all the excited states, with their respective detuning and transition
strength (see section \ref{ssec:shift-multi}).

\subsection{Optical Dipole Traps} \label{sec:trapping}

The dissipative part of the atom-light interaction is used for laser cooling of atoms \cite{Cohen-Tannoudji90}, a
pre-requisite for optical trapping. However, spontaneous photon scattering is in general detrimental to
\textit{trapped} atoms, mainly because it can induce heating and loss. Comparing equations \ref{sdip} and
\ref{sscat} yields,
\begin{eqnarray}
\frac{U_{dip}}{\hbar \gamma_{s}}=\frac{\delta }{\gamma },
\end{eqnarray}
which results in the well known fact that a trap with an
arbitrarily small scattering rate can be formed by increasing the
detuning while maintaining the ratio $I/\delta $.

Equation \ref{sdip} indicates that if the laser frequency is smaller than the resonance frequency, i.e. $\delta<0$
(``red-detuning'') the dipole potential is negative and the atoms are attracted by the light field. The minima of
the potential is found then at the position of maximum intensity. In the case $\delta>0$ (``blue detuning'') the
minima of the potential is located at the minima of the light intensity.

Trapping atoms with optical dipole potentials was first proposed by Letokhov \cite{Letokhov68} and Ashkin
\cite{Ashkin70} . Chu and coworkers \cite{Chu86}, were the first to realize such a trap, trapping about 500 atoms
for several seconds using a tightly focused red-detuned beam. Later, a far-off-resonant trap for rubidium atoms
was demonstrated \cite{Miller93}, with a detuning of up to 65 nm, i.e. $\delta>5\times 10^6 \gamma$. In this case,
the potential is nearly conservative and spontaneous scattering of photons is greatly reduced. A comprehensive
review of the different schemes and applications of such optical dipole traps is presented in reference
\cite{Grimm00}. In the limiting case where the frequency of the trapping light is much smaller then the atomic
resonance, trapping is still possible, practically with no photon scattering \cite{Takekoshi95}. Such a
quasi-electrostatic trap, formed by two crossed CO$_2$ laser beams, was used to create a Bose-Einstein condensate
without the use of magnetic traps \cite{Barrett01}.

Apart from using far-off-resonance lasers, the interaction between the light field and the atoms can be reduced by
the use of blue-detuned traps, in which atoms are confined mostly in the dark \cite{Friedman02}. Experimentally,
blue-detuned traps, which require surrounding a dark region of space with a repulsive dipole potential, are harder
to realize than red-detuned ones, where already a single focused beam constitutes a trap \cite{Chu86}. Several
configurations for far-detuned dark optical traps were demonstrated, in which gravity provided the confinement in
one direction: ``Light sheets'' traps were generated by elliptically focusing two laser beams and overlapping the
two propagating light sheets to form a ``V'' cross-section that supports against gravity, while the confinement in
the laser propagation direction is provided by the beam divergence \cite{Davidson95}. In a later work, and in
order to achieve larger trapping volume, a different trap was constructed with four light sheets producing an
inverted pyramid \cite{Lee96}. A single beam trap was demonstrated using two axicons and a spherical lens to
generate a conical hollow beam propagating upwards \cite{Ovchinnikov98}. Such gravito-optical traps are limited to
weak confinement. Traps in which light provided the confinement in all directions were developed with hollow
beams. Laguerre-Gaussian ``doughnut'' modes were used, together with additional plug-in beams, to form such a trap
\cite{Kuga97}.

Several dark traps based on a single laser beam were demonstrated, providing grater experimental simplicity and
enabling dynamical changes of the trap geometry and strength. As opposed to the 2D case, were any desired light
distribution can be generated using diffractive or refractive optical elements, there is no simple procedure to
design an arbitrary 3D light distribution \cite{Piestun94, Shabtay00}. Nevertheless, through the use of refractive
and holographic optical elements, it is possible to produce light distributions which are suitable for trapping
atoms in the dark using a single laser beam. Such light distributions, which comprise of a dark volume completely
surrounded by light, were realized using either combinations of axicons and spherical lenses, diffractive optical
elements, or rapidly scanning laser beams.

In the first scheme realized, the trapping beam was produced by passing a Gaussian beam through a phase plate of
appropriate size, which shifted the optical phase at the center of the beam by $\pi $ radians. Interference leads
to a dark volume at the focus of the lens, surrounded by light in all directions \cite{Ozeri99}. An additional
method, with a much larger volume and more symmetric shape, was realized by simultaneously focusing two
diffraction orders of a properly designed binary phase element, consisting of concentric phase rings with a $\pi
$-phase difference between subsequent rings \cite{Ozeri00}. An improved trap configuration was demonstrated by
adding an axicon telescope before the phase element of the above setup \cite{Kaplan02}. This configuration
maximizes the trap depth for a given laser power and trap dimensions, and greatly reduces the light induced
perturbations to the trapped atoms.  A related scheme was demonstrated in references \cite{Cacciapuoti01} and
\cite{Kulin01}. Finally, a tightly-focused rapidly-rotating laser beam was used to create a trap
\cite{Friedman00}. If the scan frequency is high enough, the optical dipole potential can be approximated as a
time averaged quasi-static potential. For a blue-detuned laser beam, and a radius of rotation larger than the
waist of the focussed beam, a dark volume suitable for 3D trapping is obtained.

\subsection{Photon Scattering and Coherence Relaxation in Multi-Level Atoms}
\label{ssec:back-scatter} The dipole potential can be viewed as originating from cycles of absorption of a photon
and stimulated emission into a different laser mode. This process is unfortunately accompanied by spontaneous
scattering, in which the absorption is followed by spontaneous emission in a random direction. Spontaneous
scattering is one of the main limiting mechanism for the atomic hyperfine coherence\footnote{A word of caution is
in order about our use of the terms ``decoherence'' and ``dephasing'' . Decoherence results from the dissipative
interaction of a single quantum superposition state with the environment (e.g. in the case of trapped atoms, the
coupling to the electro-magnetic vacuum which leads to spontaneous scattering of photons). In addition, the
response of a \emph{macroscopic} ensemble of quantum systems (each prepared in a superposition state) decays due
to the dephasing between the microscopic systems resulting from local variations in their evolution, and hence the
ensemble coherence is lost.}.

Spontaneous scattering is a two-photon process, in which an atom initially at a state $\ket{F,m_F}$ absorbs a
photon from the trap laser and moves to an intermediate state $\ket{F',m'_F}$ of some excited level. The atom then
decays back to the ground state, to a final state $\ket{F'',m''_F}$. If $F''=F$ and $m''_F=m_F$, the process is
called ``Rayleigh scattering''. Otherwise, it is denoted ``Raman scattering''. Clearly, Raman events destroy the
atomic state coherence and hence are detrimental to the spectroscopy. As will be discussed below in section
\ref{ssec:manypi}, also Rayleigh scattering events can be destructive for spectroscopy of a trapped ensemble.

The probability amplitude for scattering between $\ket{F,m_F}$ and $\ket{F'',m''_F}$, via an intermediate state
$\ket{F',m'_F}$ is proportional to $\braOket{F'',m''_F}{\mu_k}{F',m'_F}\braOket{F',m'_F}{\mu_j}{F,m_F}$, where
$\mu_k$ ($\mu_j$) are the spherical components of the dipole moment operator ($k,j=-1,0$ or $1$ depending on the
polarization of the absorbed and emitted photons). If the detuning from resonance is large enough such that no
specific intermediate state is resolved, then to calculate the rate of transitions $\gamma_{F,m_F \rightarrow
F'',m''_F}$ between $\ket{F,m_F}$ and $\ket{F'',m''_F}$  the amplitudes for all possible paths must be summed. For
rubidium atoms, these are the two excited states $5^2P_{1/2}$ and $5^2P_{3/2}$, with all their different hyperfine
levels and corresponding Zeeman sub-levels. We further assume that the laser polarization is linear, hence
conservation of angular momentum dictates $m_F=m'_F$. The total transition rate is \cite{Cline94}
\begin{eqnarray}\label{eq:scatt}
\gamma_{F,m_F \rightarrow F'',m''_F}=\frac{3 \pi c^2 \omega_L^3 I}{2 h \mu^4}\left| \frac{\alpha_{F,m_F
\rightarrow F''m''_F}^{(1)}}{\delta_{1}} + \frac{\alpha_{F,m_F \rightarrow F''m''_F}^{(2)}}{\delta_{2}} \right|,
\end{eqnarray}
where
\begin{eqnarray}
\alpha^{(J)}_{F,m_F \rightarrow
F'',m''_F}=\frac{\gamma_{J}}{\omega_{J}^3}\sum_{q,F',m'_F}\braOket{F'',m''_F}{\mu_q}{F',m'_F}\braOket{F',m'_F}{\mu_0}{F,m_F}.
\end{eqnarray}
Here $\omega_{J}$ and $\gamma_{J}$ are the transition frequency and excited state lifetime of the $D_{J}$ line
($J=1,2$), and $\mu = \braOket{33}{\mu_{-1}}{44}$ is the amplitude for the strongest transition.

\begin{figure}[tbp]
\begin{center}
\includegraphics[width=3in] {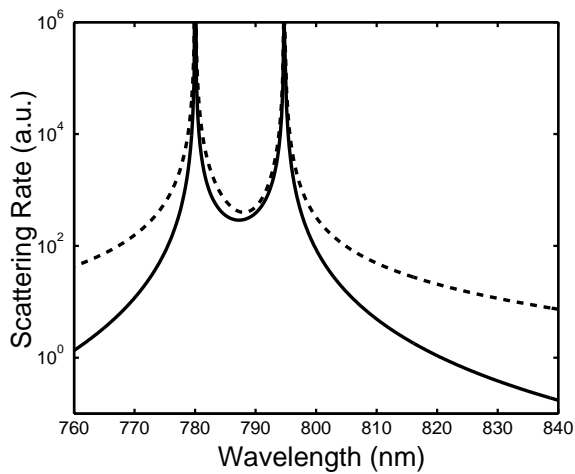}
\end{center}
\caption{Calculated total scattering rate (dashed line) and $F$-changing Raman scattering rate (full line) for
$^{85}$Rb atoms and linear polarization. The former affects the coherence time of trapped atoms, but the later is
experimentally simpler to measure.} \label{scattering}
\end{figure}

The first step in spectroscopy of trapped $^{85}$Rb atoms, is to transfer the entire population to $F=2$. There,
it is distributed uniformly between $m_F$ levels. We then apply a certain microwave pulse sequence (e.g. a Rabi
pulse) which acts only on the transition $\ket{F=2,m_F=0} \rightarrow \ket{F=3,m_F=0}$, since a bias magnetic
field is applied to shift all other transitions from resonance. Finally, we detect the total population of $F=3$.
We would like to measure, independently, the rate of incoherent scattering events that also affect the population
of $F=3$, in order to subtract them from our coherently driven population changes. Experimentally, we can measure
the amount of $F$-changing transitions, but not the rate of Rayleigh scattering events, or Raman events that do
not change $F$ \cite{Cline94}. Equation \ref{eq:scatt}, together with the appropriate summation, is used to
calculate the relative amount of different scattering processes. For example, the rate $\gamma_{F=2 \rightarrow
F=3}$ for a transition between hyperfine levels $2$ and $3$ is calculated by averaging $\gamma_{F,m_F \rightarrow
F'',m''_F}$ over initial $m_F$ levels of $F=2$, and summing over final $m''_F$ levels of $F=3$. The total
scattering rate is given by summing over $m_F$ levels of $F=2$ and $m''_F$ levels of $F=3$. Figure
\ref{scattering} shows a calculation of the relative amount of these two processes, as a function of the laser
wavelength. Using the results of figure \ref{scattering} we calculate the total amount of scattering events out of
our measurement.

Note, that for a detuning larger than the fine-structure splitting of the excited state ($15 \nm$ in rubidium), a
destructive interference exists between the transition amplitudes for Raman scattering, summed over the
intermediate excited states \cite{Cline94}. In this case, most spontaneous scattering events leave the internal
state of the atom unchanged.

\subsection{Light Shifts and Dephasing in Multi-Level Atoms}
\label{ssec:shift-multi} For a real multi-level atom, the light-shifts described in section \ref{sec:shift}
depend, in general, on the particular substate of the atom. The equivalent of equation \ref{shift2} in the case of
a multi-level atom can be written using the dipole matrix elements $\mu_{ij}=\braOket{e_i}{\mu}{g_i}$ between a
specific ground state $\ket{g_i}$ and a specific excited state $\ket{e_j}$. Using
$\mu_{ij}=c_{ij}\,\left\|\mu\right\|$, we can write the ac Stark shift for a ground state $\left|
g_{i}\right\rangle $ as \cite{Grimm00}:
\begin{eqnarray}
\Delta E_{i}=\frac{3\pi c^{2}\gamma}{2\omega_{0}^{3}}I \times \sum \frac{c_{ij}^{2}}{\delta_{ij}},
\label{eq_lightshift}
\end{eqnarray}
The summation takes into account the contributions of the different coupled excited levels $\ket{e_{j}}$, each
with its respective transition coefficient $c_{ij}$, and detuning $\delta_{ij}=\omega- \omega_{ij} $.

If the detuning of the light is large as compared to the excited state hyperfine splitting $\ohf^{\prime}$, then
certain sum rules exist for the transition coefficients \cite{Deutsch98} and  equation \ref{eq_lightshift} for the
shift of a ground state with total angular momentum $F$, simplifies to:
\begin{eqnarray}
\Delta E_F =\frac{ \pi c^{2}\gamma I}{2 \omega_{0}^{3}} \left[ \left(
\frac{1}{\delta_{1,F}}+\frac{2}{\delta_{2,F}}\right)- g_F m_F \sqrt{1-\epsilon^2} \left(
\frac{1}{\delta_{1,F}}-\frac{1}{\delta_{2,F}}\right)\right],
\end{eqnarray}
where $\epsilon$ is the light polarization ellipticity  and $\delta_{1,F}$ [$\delta_{2,F}$] is the detuning of the
laser from the $D_1$ [$D_2$] line. For linearly polarized light ($\epsilon=1$)
\begin{eqnarray}
\Delta E_F\left( \mathbf{r}\right) =\frac{3 \pi c^{2}\gamma}{2
\omega_{0}^{3}} \frac{I\left( \mathbf{r}\right) }{\delta^*_{F}},
\label{deltaE}
\end{eqnarray}
where
\begin{eqnarray}
\frac{1}{\delta^*_{F}}= \frac{1}{3} \left[ \frac{2}{\delta_{2,F}} + \frac{1}{\delta_{2,F}-\Delta_F}\right]
\end{eqnarray}
is the ``weighted detuning'' (below, we drop the ``$*$'' and call this the detuning). Note, that the interaction
matrix elements are identical for atoms in $\ket{F=2}$ and $\ket{F=3}$. However, the light shift is inversely
proportional to the trap laser detuning, which differs by $ \ohf$, and therefore the potential from a far-detuned
laser is slightly different for the two hyperfine levels. Loosely speaking, this difference in potential means
that different hyperfine states ``feel'' different trap shapes. For microwave spectroscopy of a thermal ensemble
this results in a rapid decay of the ensemble-averaged spectroscopic signal as a consequence of dephasing, much
faster than the decay due to spontaneous scattering of photons.

\section{Experimental Setup} \label{ap:setup}

All the experiments described in this tutorial consist of three stages. During the first stage standard
laser-cooling techniques are used to cool an ensemble of $^{85}$Rb atoms and load a portion of them into a
far-off-resonance trap. During the second stage, all the nearly-resonant laser beams are shut-off leaving only the
far-off-resonance trapping beam on, and microwave pulses are applied. In last stage short pulses of on-resonance
beams are used for the diagnostics. In this section, a detailed description of the above steps is provided.

\subsection{Laser Cooling and Trapping}
\begin{figure}[tbp]
\begin{center}
\includegraphics[width=5in] {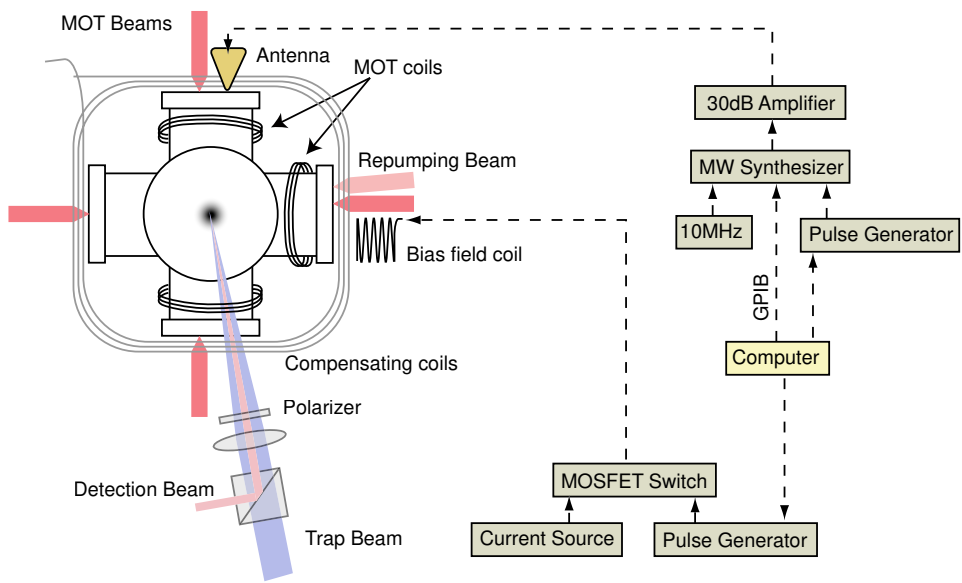} \caption{Basic components of the experimental setup.
The MOT beams consists of three orthogonal pairs of counter-propagating beams, and a repumping beam. Two copper
coils in the anti-Helmholtz configuration provide the magnetic field gradient, and three orthogonal Helmholtz
coils are used to compensate for constant magnetic fields. The atoms are imaged with an intensified CCD camera,
and their fluorescence measured with a photomultiplier tube. The dipole trap beam enters the chamber,
co-propagating with the detection beam. The microwave pulses are created by a synthesizer, amplified and radiated
into the chamber using an antenna. A bias magnetic field is applied parallel to the trap's polarization axis and
to the microwave magnetic field direction, in order to Zeeman shift the magnetic sensitive $m_{F}\neq 0$ levels
out of resonance with the microwave pulse. } \label{setup_basic_all}
\end{center}
\end{figure}

The heart of our setup is a vacuum chamber connected to a small reservoir of rubidium atoms through a valve. The
chamber has six big windows used for the laser beams, and two additional small windows used for imaging the atomic
cloud with an intensified CCD camera and measuring the fluorescence signal with a photomultiplier tube (see figure
\ref{setup_basic_all}).

The first step in all the experiments described here is a magneto-optical trap (MOT) \cite{Raab87}. The MOT laser
beams consist of three orthogonal pairs of counter-propagating beams, detuned approximately $-17 \Mhz$
($=-2.8\gamma$) from the $\ket{5S_{1/2},F=3} \rightarrow \ket{5P_{3/2},F'=4}$ line, and a ``repumping'' beam in
resonance with $\ket{5S_{1/2},F=2} \rightarrow \ket{5P_{3/2},F'=2}$. A pair of water-cooled copper coils in the
anti-Helmholtz configuration provide the magnetic field gradient for the MOT (see figure \ref{setup_basic_all}).
In addition, three orthogonal Helmholtz coils are used to compensate for constant magnetic fields.

The MOT loading time is $\sim 700$ ms. After that a $\sim 50$ ms temporal dark MOT stage \cite{Adams95} is applied
in order to increase the spatial density of the atoms. The intensity of the MOT beams is reduced to $\sim 1.5
\mw\cm^{-2}$, their detuning increased to $\sim -30 \Mhz$, the magnetic field gradient increased to $8
\g\cm^{-1}$, and the repump intensity reduced by a factor of 40. Final optimization of these parameters is
performed by optimizing the number of atoms loaded into the dark optical trap. A peak density of $\sim 1\times
10^{11}$ cm$^{-3}$ is achieved for optimized parameters, but typically we work at lower densities, for which
collisions are negligible for the timescales of our experiments. The laser cooling stage ends with a
polarization-gradient cooling stage \cite{Dalibard89}, in which the magnetic field is set to zero, the MOT beams
intensity are further decreased and their detuning increased. Usually, $3 \msec$ of this cooling stage result in a
temperature of $\sim 10-20 \mck $.

The dipole trap in our experiments consists of a linearly polarized horizontal Gaussian laser beam with wavelength
in the range $\lambda =785-810 \nm$. The trap beam is coupled into a single mode fiber, and the output of the
fiber (with power in the range $P=50-400 \mw$) is focused to a $w_0\sim 50\mcm$ spot at the center of the vacuum
chamber. An active servo circuit and an acousto-optic modulator ensure a $1\%$ stability in the beam power, and
hence in the trap depth. With typical values of $P=50 \mw$ and $\lambda =800 \nm$ we achieve a trap with a depth
of $\sim 35 \mck$, and oscillations frequencies $\omega_r=2.3 \khz$ and $\omega_z=8.4 \hz$ in the radial and axial
dimension, respectively. The clear separation of time-scales between the fast transverse oscillations and the very
slow longitudinal ones is used in the analysis of some of our experiments to neglect the longitudinal motion,
which is essentially frozen during the duration of the experiment, and hence to treat the system as a
two-dimensional one.

The trap beam overlaps the center of the atomic cloud during the cooling stages and, after turning off all the
lasers (with the exception of the trap) we end up with $\sim 10^5$ atoms loaded in the trap (depending on the
specific power and detuning) with a temperature of $\sim 20 \mck$. At the end of this stage, the atoms are
prepared in the $F=2$ ground state by turning on the MOT beams, without a repump beam \cite{Ozeri99}, for $1
\msec$.

\subsection{Microwave Spectroscopy}
We are interested in spectroscopy of the ``clock'' transition, i.e. the $\ohf=2\pi \times 3.032 \times 10^9
\sec^{-1}$ transition between the two magnetic insensitive hyperfine Zeeman substates of the ground state,
$\ket{F=2,m_F=0}$ and $\ket{F=3,m_F=0}$, which we denote $\ket{\downarrow}$ and $\ket{\uparrow}$. We drive this
transition with a nearly-resonant microwave field. The microwave pulses are created by a synthesizer (Anritsu,
69317B), whose clock is connected to a high stability and low phase noise 10 MHz oscillator. The pulses are then
amplified with a 30 dB amplifier (Mini-Circuits, ZVE-86) and irradiated into the chamber using a Log-periodic
antenna. The strongest microwave fields produced in our setup correspond to a Rabi-frequency of 5 kHz for free
(untrapped) atoms.

A bias magnetic field is applied parallel to the trap's polarization axis and to the microwave magnetic field
direction, in order to Zeeman shift the magnetic sensitive $m_{F}\neq 0$ levels out of resonance with the
microwave pulse. In most of the experiments (See sections \ref{sec:compensating} and \ref{sec:echo}), its value is
$\sim 40-80 \mg$. For the experiment in section \ref{ssec:manypi}, it is $\sim 240 \mg$, but then a MOSFET switch
is used to turn it on after the cooling stage, which requires a nearly zero field. Special care is taken to align
the directions of the microwave magnetic field and the bias magnetic field to be parallel to the trap's laser
magnetic field (i.e. the trap's polarization is perpendicular to the bias field), to enable a common well-defined
quantization axis for all the fields interacting with the atoms.

\subsection{Diagnostics}
Following the microwave pulses, $N_{3}$ (the population in $F=3$) is measured by detecting the fluorescence during
a short pulse of a laser beam resonant with the cycling transition $\ket{5S_{1/2},F=3}\rightarrow
\ket{5P_{3/2},F=4}$. The population of $F=2$ is then measured for the same experimental run by turning on the
repumping beam (which is resonant with $\ket{5S_{1/2},F=2}\rightarrow \ket{5P_{3/2},F=3}$) and applying an
additional detection pulse. This normalized detection scheme is insensitive to shot-to-shot fluctuations in atom
number as well as fluctuations of the detection laser frequency and intensity \cite{Khaykovich00}.

In addition to the population of $\ket{\ua}$ by the microwave field, $\ket{\ua}$ and all sublevels of $F=3$ are in
general populated by spontaneous scattering of photons from the trapping laser, which always accompanies the
dipole potential and tends to destroy the atomic coherence (see section \ref{ch:back}). A simple way to measure
the amount of photon scattering is by measuring the spin relaxation caused by spontaneous Raman scattering
\cite{Cline94,Ozeri99}. This is a very useful experimental technique which enables to measure even very low
scattering rates. For this measurement, the trapped atoms are first prepared in the lower hyperfine level of the
ground state, $F=2$. The number of atoms in $F=3$ after a variable time $t$, $N_{3}\left( t\right) $, is measured
by detecting the fluorescence after a short pulse of a resonant laser beam. The fraction of Raman scattering
events out of the total photon scattering events can be calculated using the known matrix elements for the
transition, and therefore the total amount of photon scattering can be inferred from the spin relaxation
measurement (see section \ref{ssec:back-scatter}).

We calculate the microwave driven part of the population by subtracting from the measured signal the above
contribution due to $F$-changing Raman transitions induced by the trap laser and normalize to the signal after a
short $ \pi $-pulse, which transfers the whole population of $\ket{\da}$ to $\ket{\ua}$. The corrected and
normalized signal is denoted $P_{\ua}$.

\section{Microwave Spectroscopy in a Dipole Trap: A Classical Model}
\label{sec:classical}

\begin{figure}[tbp]
\begin{center}
\includegraphics[width=3.6in]{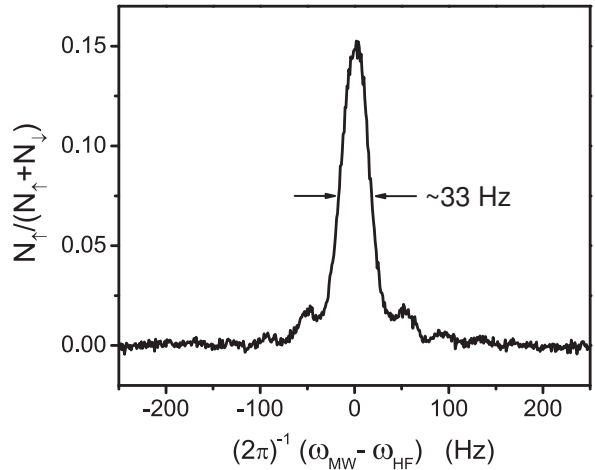}
\end{center}
\caption{Rabi spectrum of free atoms with a $25 \msec$ microwave pulse, showing the characteristic ``sinc$^2$''
lineshape corresponding to a rectangular pulse. Note that since the population of the four $\ket{F=2,m_F\neq0}$
states is included in $N_{\ua}/(N_{\da}+N_{\ua})$, a value of 0.2 represents the maximal possible signal (a $\pi$
pulse) for the $\ket{F=2, m_F=0}\rightarrow \ket{F=3, m_F=0}$ transition.} \label{freerabi}
\end{figure}

We drive the magnetic-insensitive transition with a nearly-resonant microwave field, and apply a static magnetic
field in order to shift all other (magnetic sensitive) levels out of resonance. In such case, and neglecting for
the moment the effects of the trap on the atoms, we can consider a two-level system, separated in energy by
$\ehf=\hbar \ohf$. When a free (untrapped) atom, initially in state $\ket{\downarrow}$, is placed in a microwave
field with frequency $\omw$ the probability of finding it in state $\ket{\uparrow}$ as a function of time is:
\begin{eqnarray}
P_\uparrow=\frac{\Omw^2}{\Omw'^2}\sin^2\left(\frac{\Omw'}{2}t \right),\label{eq:rabi_prob}
\end{eqnarray}
where $\Omw$ is the Rabi frequency of the microwave field, $\Omw'=\sqrt{\Omw^2+\Delta^2}$ is the generalized Rabi
frequency, $\Delta = \omw - \ohf$ is the microwave detuning, and we have used the rotating wave approximation. As
a function of the interaction time, an atom oscillates between the two states at the effective Rabi frequency,
where the contrast of the oscillation is determined by the Rabi frequency and the detuning. This sinusoidal
population transfer is referred to as Rabi flopping. If the interaction time is fixed and the power is varied,
$P_\uparrow$ reaches a maximum for $\Omw t=\pi$. For such a pulse, the transition probability as a function of
frequency has a maximum for $\omw=\ohf$, and (FWHM) width of $\Delta \omega / 2 \pi \approx 0.8/t$. A measurement
of the resonance frequency using a single pulse of constant duration is called Rabi spectroscopy. Figure
\ref{freerabi} shows a Rabi spectrum of cold atoms freely falling after the shut off of the cooling and trapping
beams. The pulse duration is limited by the falling time of the atoms from the detection volume. For this short
pulse we observe a Fourier-limited linewidth, indicating that no decoherence or dephasing mechanisms exist for for
free-falling atoms in this time-scales.

An alternative and widely used method is Ramsey spectroscopy, in which two consecutive pulses of duration $t$,
separated by a ``dark'' period of duration $T$, are used. The first pulse (called a $\frac{\pi}{2}$ pulse) puts
the atoms in a quantum superposition of both states. During free evolution, due to the detuning of the driving
field from resonance, these two states develop a relative phase in the rotating frame of the field. The second
$\frac{\pi}{2}$ pulse causes the two states to interfere, thus translating phase differences into population
differences. If the field frequency is scanned, the phase difference acquired in the dark period, and thus also
the population of the levels after the second pulse, oscillate. Following a two-pulse Ramsey sequence, the
probability of being in the excited state as a function of time for $\abs{\Delta} \ll \Omw$ and $t \ll T$ is
simply
\begin{eqnarray}
P_\uparrow=\frac{1}{2}+\frac{1}{2}\cos\left(\Delta T\right).
\end{eqnarray}

When performing Rabi or Ramsey Spectroscopy on optically trapped atoms, the above treatment no longer holds. The
dipole potential is inversely proportional to the detuning of the trapping laser beam from resonance. Since there
is a slight difference in potential for atoms in different hyperfine states, the external (center of mass)
potential depends on the internal (spin) state, hence the internal and external degrees of freedom cannot be
separated.

We first consider a simple model of Rabi spectroscopy, in which we assume the atoms to be in thermal equilibrium
in the potential of the trap and frozen in position during the microwave pulse. In order for this model to be
valid, the microwave pulse duration $t$ must be much shorter than the typical oscilation time in the trap, i.e.
\begin{eqnarray}
t \ll \omega_{osc}^{-1}.\label{firstcon}
\end{eqnarray}

Let us look now at the ground states of atoms trapped in a dipole trap. From equation \ref{deltaE} it is seen that
in the presence of the trap light (and assuming for simplicity that $\delta_{2,F} \ll \Delta_F$, so the
contribution from the $D_1$ line is neglected) the energy splitting is modified by
\begin{eqnarray}
\hbar\widetilde{\ohf}\left( \mathbf{r}\right) -\hbar\ohf= \frac{\pi
c^{2}\gamma}{\omega_{0}^{3}}\frac{\ohf}{\delta^{2}}\left[ \frac{1}{1-\left( \frac{ \ohf}{2\delta}\right)
^{2}}\right] I\left( \mathbf{r}\right), \label{eq_newhf}
\end{eqnarray}
where $\widetilde{\ohf}\left( \mathbf{r}\right) $ is the spatially dependent transition frequency in the presence
of the light and $\delta \triangleq(\delta_{2,F=2}+\delta_{2,F=3})/2$ measures the laser detuning from the center
of the ground state hyperfine splitting. Equation \ref{eq_newhf} indicates that for $\left| \delta\right| >\frac{
\ohf}{2}$, i.e. for $ \delta_{2,F=2} $ and $\delta_{2,F=3}$ both positive (or both negative), the ground state
energy splitting is always reduced by the presence of a light field. When the detuning is ``between'' the
hyperfine levels, $\left| \delta\right| <\frac{\ohf}{2}$, the energy splitting is enlarged.

For a trap with detuning $ \delta \gg\frac{\ohf}{2}$ equation \ref{eq_newhf} yields
\begin{eqnarray}
\widetilde{\ohf}\left( \mathbf{r}\right) -\ohf\approx\left( \frac{\ohf}{ \delta}\right) \frac{U\!\left(
\mathbf{r}\right)}{\hbar},
\end{eqnarray}
where $U\!\left( \mathbf{r}\right) $ is the spatially dependent dipole potential that forms the trap. The fact
that the relative ac Stark shift is $\ohf\diagup\delta$ smaller than the dipole potential, is the main motivation
for using far-off-resonance traps for precision spectroscopy. For example in reference \cite{Davidson95} the
relative ac Stark shifts were only $\ohf\diagup\delta\approx 2\cdot10^{-4}$ times the dipole potential.

For a trapped atomic ensemble, averaging over the different values of $U\!\left( \mathbf{r}\right) $ causes a
shift in the ensemble averaged transition frequency , $\left\langle\widetilde{\ohf}-\ohf\right\rangle$. This shift
is also a function of the trap's geometry, which determines the degree of exposure of the atoms to the trapping
light. Following references \cite{Grimm00} and \cite{Friedman02}, we introduce the parameter $\kappa$, defined as
the ratio of the ensemble-averaged potential and kinetic energies of the trapped atoms, $\kappa =\left\langle
U\right\rangle /\left\langle E_{k}\right\rangle $, and refer to it as the ``darkness factor'' of the trap.
Assuming a trapped atomic gas in thermal equilibrium, the ensemble-averaged kinetic energy is $\left\langle
E_{k}\right\rangle =\frac{3}{2}k_{B}T$, and neglecting gravity, the ensemble averaged potential energy is given by
\begin{eqnarray}
\left\langle U\right\rangle = {\displaystyle{\frac{\displaystyle\int d \mathbf{r}\,U(\mathbf{r})\exp \left[
-\frac{U(\mathbf{r })-U_{0}}{k_BT}\right] }{\displaystyle\int d\mathbf{r}\,\exp \left[ -
\frac{U(\mathbf{r})-U_{0}}{k_BT}\right] }}},
\end{eqnarray}
where $ U_{0} $ is the potential at the trap bottom and the integration is over the entire trap
volume\footnote{For red-detuned traps, $U_{0}<0$, while $U_{0}=0$ for most blue-detuned traps. For example, for an
harmonic trap, the ensemble-averaged potential energy (relative to the trap's bottom) and the averaged kinetic
energy are equal, and therefore the darkness factor is always $\kappa=1$ for a blue-detuned trap, independent of
laser power, detuning or atomic temperature, but $\kappa\propto U_{0}/k_{B}T$ for a red-detuned trap with
$U_{0}>>k_{B}T$.}. The ensemble frequency shift is then given by
\begin{eqnarray}
\left\langle\widetilde{\ohf}-\ohf\right\rangle =\frac{\ohf}{\hbar \delta} \cdot \frac{3}{2}k_{b}T\cdot \kappa .
\end{eqnarray}
Using equation \ref{sscat} we can calculate the average spontaneous scattering rate, which determines the
\emph{homogeneous} coherence time of the atomic ensemble:
\begin{eqnarray}
\left\langle \gamma _{s}\right\rangle =\frac{\gamma}{\hbar \delta}\cdot \frac{3}{2} k_{b}T \cdot \kappa
.\label{eq:sr}
\end{eqnarray}
In addition, the spatial dependence of $ U\!\left( \mathbf{r}\right) $ will result in an inhomogeneous broadening
responsible for the \emph{inhomogeneous} decoherence time:
\begin{eqnarray}
\sigma (\ohf) \equiv\sqrt{\left\langle\widetilde{\ohf}^{2}\right\rangle-
\left\langle\widetilde{\ohf}\right\rangle^{2}} = \frac{\ohf}{\hbar \delta} \sqrt{\left\langle U\!\left(
\mathbf{r}\right)^{2}\right\rangle- \left\langle U\!\left( \mathbf{r}\right) \right\rangle^{2}}.\label{eq:brod}
\end{eqnarray}

For example, for a thermal ensemble in an harmonic trap, we have
\begin{eqnarray}
\left\langle \gamma _{s}\right\rangle =\frac{\gamma}{\hbar \delta} \left(\frac{3}{2} k_{b}T \right),\label{eq:msr}
\end{eqnarray}
and
\begin{eqnarray}
\sigma (\ohf)=\left( \frac{\ohf}{\delta}\right) \frac{1}{\hbar} \sqrt{\frac{3}{2}}k_BT.\label{eq:class_broad}
\end{eqnarray}
Using equations \ref{eq:msr} and \ref{eq:class_broad} we can see that
\begin{eqnarray}
\frac{\sigma (\ohf)}{\left\langle \gamma _{s}\right\rangle} \sim \frac{\ohf}{\gamma}.\label{nir}
\end{eqnarray}
For most alkali-metal atoms, such as rubidium and cesium, ${\ohf}/{\gamma}\sim10^3$, indicating that, however
small, the relative ac Stark broadening is still much larger than the spontaneous photon scattering rate. The
homogeneous and inhomogeneous coherence times are inversely proportional to the spontaneous scattering rate and
the ac Stark broadening, respectively, hence equation \ref{nir} shows that the latter is the main limiting factor
for the atomic coherence time in the trap.

Note, that in order to resolve the frequency distribution, the Fourier broadening must be smaller than the
inhomogeneous broadening, i.e. $t^{-1} \ll \sigma(\ohf)$. Hence, using also equation \ref{firstcon}, we conclude
that the requirement for the present classical ``stationary'' model to be both valid and ``interesting'' is
\begin{eqnarray}
\omega_{osc} \ll \sigma(\ohf).
\end{eqnarray}

\section{Case Study: Inhomogeneous Broadening and Trap Geometry}
\label{sec:case}

As a specific example for the use of the classical model presented in the previous section we compare the
performance of four different trap geometries. One trap is red-detuned (the crossed Gaussian beams trap
\cite{Adams95}), and three are blue-detuned: the Rotating Beam Trap (ROBOT) \cite{Friedman00} is based on a
repulsive optical potential formed by a tightly focused blue-detuned laser beam which is a rapidly scanned using
two perpendicular acousto-optic scanners. The Laguerre-Gaussian (LG) trap \cite{Kuga97}, consists of a hollow
laser beam (LG$_{0}^{3}$) and two additional ``plug'' beams that confine the atoms in the propagation direction of
the hollow beam. Finally, the``optimal trap'' \cite{Kaplan02} uses two refractive axicons and a binary phase
element to create a spatial light distribution consisting of two hollow cones attached at their bases and
completely surrounding a dark region.

\begin{figure}[tbp]
\begin{center}
\includegraphics[width=3.15in]{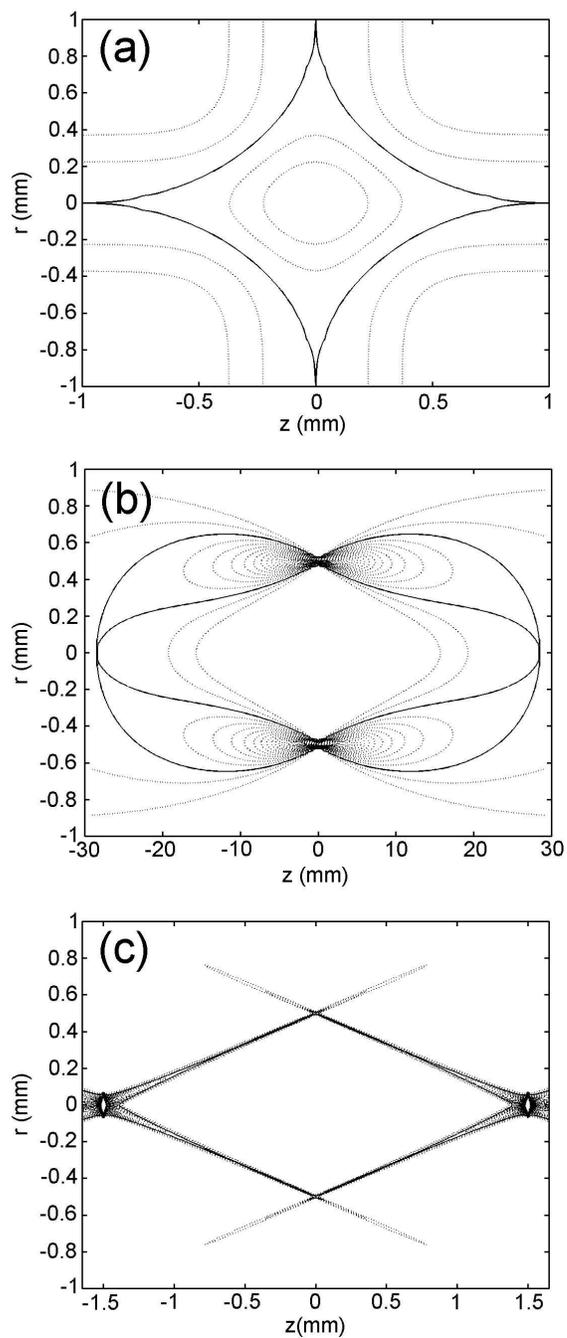}
\end{center}
\caption{Contour maps of the calculated trapping depth for three different optical traps. The dashed lines are
equidistant contours of equal potential. The solid line is the contour corresponding to the trap depth. All the
traps have the same radial dimension. (a) Crossed Gaussian beam (red-detuned) trap. (b) Rotating beam trap. (c)
``Optimal'' trap.} \label{trap_contours}
\end{figure}

Figure \ref{trap_contours} shows the calculated optical potential distribution of three of these traps, in the
$r-z$ plane. The dashed lines in figure \ref{trap_contours}\ are equidistant contours of equal potential, and the
solid one is the contour line corresponding to the trap depth. We assume a trap laser with a fixed power $P=1W$
and a sample of $^{85}$Rb atoms, which are laser cooled to a temperature of $5 \mck$, and form a nearly spherical
cloud with a radius of $0.5$ mm, typical parameters for a magneto-optical trap.

Adopting a criteria of $>90\%$ geometrical loading efficiency from the magneto-optical trap, we choose a radius
$r=0.5$ mm for all dark traps. The beam waist is chosen as $w_{0}=50 \mcm$ for the ROBOT, and $ w_{0}=10\mcm$ for
the optimal trap. The length of the latter is an independent parameter, chosen as $L=3$ mm to optimize the power
distribution as explained in reference \cite{Kaplan02}. We neglect the enhanced loading efficiency of red-detuned
traps \cite{Kuppens00}, and assume for the crossed trap $w_{0}=0.6$ mm, which corresponds to $>90\%$ overlap with
the magneto-optical trap \footnote{We choose a crossed trap, and not a simpler focused gaussian beam trap, since
with a single focused beam a trap radius of 0.6 mm will result in an extremely large axial size of $>1$ m.}
\cite{Adams95}.

The detuning in the comparison is chosen such that the depth of each trap is $3$ times larger than the mean
kinetic energy of the atoms. Since a fixed laser power is assumed, less efficient traps would require a smaller
detuning to provide the same trap depth. In table \ref{tb:optimal}, the calculated required detuning in each of
the different traps, with the parameters discussed above, is presented together with the calculated darkness
factor $\kappa$. As expected, all blue-detuned traps have a better darkness factor than the red-detuned trap. The
optimal trap has a significantly better darkness factor ($\kappa =0.02$) than all other schemes, due to its very
thin walls and nearly minimal surface to volume ratio. Next, the mean spontaneous photon scattering rate
$\left\langle \gamma _{s}\right\rangle$, and the inhomogeneous hyperfine frequency broadening $\sigma(\ohf)$ are
calculated using equations \ref{eq:sr} and \ref{eq:brod}, respectively. Here, the advantage of the optimal trap is
even much larger, since the improved darkness factor is combined with the efficient distribution of optical power
that enables an increased detuning for the same trap depth. For example, the inhomogeneous broadening in the
optimal trap is only $47 \hz$, while for all the others is of the order of a few $\khz$.

Including gravity in our calculations, results in an increase of 10-60\% for the scattering rate and inhomogeneous
broadening for rubidium atoms in the above dark traps. For the lighter alkali-metal atoms (e.g. sodium and
lithium), the inclusion of gravity yields an increase of no more than 5\% in the scattering rate.

Our classical model and the above comparison show that, in addition to the use of far-off-resonance lasers and
dark optical traps, geometry can also be exploited to minimize the undesired exposure of the atoms to the trapping
light.

\begin{table}[tbp]\label{tb:optimal}
\begin{tabular}{|l|l|l|l|l|}
\hline & $\delta$ (nm) & $\kappa $ & $\left\langle \gamma _{s}\right\rangle$ (s$^{-1}$) & $\sigma(\ohf)$ (Hz)\\
\hline Red
detuned trap & -0.7 & 4.9 & 166.9 & 1.9$\cdot$10$^{3}$\\ \hline LG trap & 0.23 & 0.6 & 87.6 & 5$\cdot$10$^{3}$\\
\hline ROBOT & 0.19 & 0.2 & 21.2 & 3.8$\cdot$10$^{3}$\\ \hline ``Optimal'' trap & 4.69 & 0.02 & 0.09 & 47.2\\
\hline
\end{tabular}
\caption{Required detuning, calculated atomic darkness factor, mean spontaneous photon scattering rate and
inhomogeneous frequency broadening of the hyperfine splitting for $^{85}$Rb atoms confined in the traps analyzed
in the text.}
\end{table}

\section{A "Compensated" Trap}\label{sec:compensating}
The previous section indicated that even for far detuned traps with favorable geometries the hyperfine coherence
of optically trapped atoms is still predominately limited by the difference in the trap-induced ac Stark shifts
between the two hyperfine levels. To cancel these relative ac Stark shifts, we introduce an additional laser beam,
with intensity $I'\left( \mathbf{r}\right)$ and frequency between the resonant frequencies of the two ground state
hyperfine levels, say in the middle, i.e. $-\ohf/2$ and $+\ohf/2$ from the lower and higher hyperfine level
respectively\footnote{The detuning of the compensating laser can be chosen in the range
$-\frac{\ohf}{2}<\delta<\frac{\ohf}{2}$, yielding a straightforward modification of equations
\ref{Newhfs},\ref{Etha}; however, choosing $\delta=0$ minimizes spontaneous photon scattering.}. The total shift
is obtained by adding the shifts from the trap and the ``compensating'' beam,
\begin{eqnarray}
\widetilde{\ohf}\left( \mathbf{r}\right) -\ohf=\frac{\pi c^{2}\gamma\ohf}{ \omega_{0}^{3}}\left[ \frac{I\left(
\mathbf{r}\right) }{\delta^{2}-\left( \frac {\ohf}{2}\right) ^{2}}-\frac{I'\left( \mathbf{r}\right) }{\left(
\frac{\ohf}{2}\right) ^{2}} \right] \label{Newhfs}.
\end{eqnarray}
If the compensating beam is spatially mode-matched with the trap beam, i.e. $ I'\left( \mathbf{r}\right)
=\eta\times I\left( \mathbf{r}\right) $, then a complete cancellation of the inhomogeneous broadening will occur
for
\begin{eqnarray}
\eta= \frac{(\frac{\ohf}{2})^{2}}{\delta^{2}-\left( \frac{\ohf}{2}\right) ^{2}}\approx(\frac{\ohf}{2\delta})^{2}.
\label{Etha}
\end{eqnarray}

\begin{figure}[tbp]
\begin{center}
\includegraphics[width=4in]{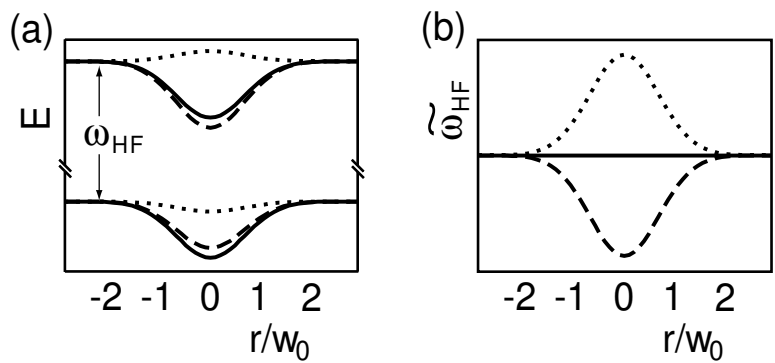}
\end{center}
\caption{Ground level energies (a) and energy difference (b) of atoms trapped in a focused Gaussian beam. When
exposed to the trapping laser, the two hyperfine levels have a different ac Stark shift (dashed line). An
additional weak laser, detuned to the middle of the hyperfine splitting, creates an ac Stark shift (dotted line)
such that the total amount of light shift (full line) is identical for both hyperfine levels.} \label{complevels}
\end{figure}

Figure \ref{complevels} shows the shift of the hyperfine levels (a) and  the hyperfine energy difference (b)
caused by the trapping beam (dashed line) and compensating beam (dotted line). In the presence of both beams, the
levels are shifted by the same amount (full line). As a specific example, with a trap detuned by $5 \nm$ we have
$\eta\approx3.6\times10^{-7}$. Hence, with a typical trap power of $50 \mw$, the required compensating beam power
is $20 \nw$. (A similar calculation can be made which takes into account also the contribution of the D$_1$
transition, and introduces only a small correction to equation \ref{Etha}).

Note, that the dipole potential created by the compensating beam is $U'\left( \mathbf{r}\right) =\pm
\frac{1}{2}\frac{\ohf}{\delta}U\left( \mathbf{r} \right) $ for atoms in the upper an lower hyperfine level,
respectively, and hence is negligible when compared with the potential of the dipole trap. Moreover, the photon
scattering rate from the compensating beam $\gamma'_{s}$ is given by $\hbar \gamma'_{s}\approx2\gamma U'/\ohf$,
which can be also written as $\hbar \gamma'_{s}\approx\frac{\gamma }{\delta}U\left(
\mathbf{r}\right)\approx\hbar\gamma_{s}$. Hence, the scattering rate from the nearly resonant compensating beam,
is of similar magnitude to that of the far-off-resonance trapping beam, $\gamma_{s}$.

We implement the proposed scheme with a red-detuned ($\lambda=785 \nm$) Gaussian trap \cite{Chu86}, created by
focusing a $50 \mw$ laser to a waist of $w_{0}=50\mcm$ (See section \ref{ap:setup}) and resulting in a potential
depth of $U_{0}\approx 34 \mck$ and oscillation frequencies of $ 2300$ and $8 \hz$ in the radial and axial
directions, respectively. An additional laser, with frequency locked close to the middle of the ground state
hyperfine splitting, is combined with the trap laser \cite{Kaplan02b}. To achieve optimal spatial mode-match, both
lasers are coupled into a polarization-preserving single-mode optical fiber, and the fiber's output is passed
through a polarizer and focused into the vacuum chamber. Two servo loops are used to control and stabilize the
power of the lasers: The first one ensures a $1\%$ stability of the trap laser. More importantly, for complete
compensation of the relative ac Stark shifts, a second servo loop ensures a $0.1\%$ stability of the power ratio
$\eta$ throughout the entire duration of the experiment. Since typically $\eta\sim10^{-7}$ in our experiment, the
beams are separated by two gratings and two pinholes before their power can be measured independently. The loading
procedure is described in section \ref{ap:setup}.

\begin{figure}[tbp]
\begin{center}
\includegraphics[width=3.75in]{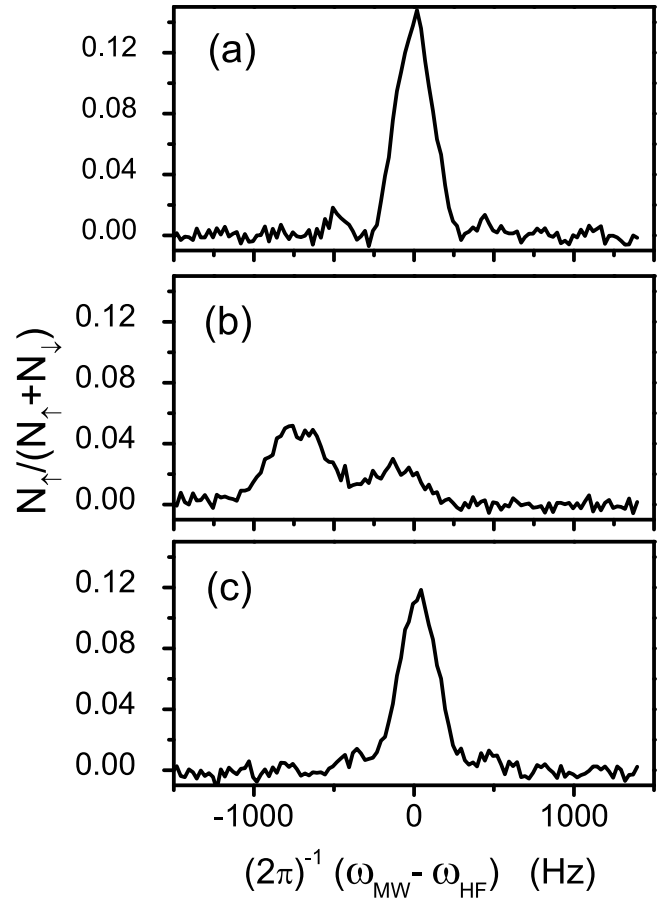}
\end{center}
\caption{Rabi spectrum of the hyperfine splitting of $^{85}$Rb,
with a $3 \msec$ pulse. (a) Spectrum of free falling atoms. (b)
Spectrum of trapped atoms, showing a shift in the line center and
a broadening. (c) Spectrum of trapped atoms, with an additional
compensating beam. The addition of the weak compensating beam,
nearly cancels the shift and broadening of the spectrum.}
\label{spectrum3}
\end{figure}

Figure \ref{spectrum3} shows results for the Rabi spectrum with a $3 \msec$ long $\pi$ pulse. A constant
background resulting from spontaneous $F$-changing Raman scattering \cite{Cline94} is subtracted. The spectrum of
free-falling atoms (figure \ref{spectrum3}a) shows no inhomogeneous broadening and a RMS width, $\sigma$, which is
Fourier limited to $110 \hz$. A shift in the peak frequency ($-756 \hz$), and a broadening of the line (to
$\sigma=320 \hz$) are seen in the spectrum of trapped atoms (figure \ref{spectrum3}b), in fair agreement with the
calculated trap depth and atomic temperature. This inhomogeneous broadening is not significantly affected by the
duration of the pulse. The addition of the weak compensating beam, nearly cancels the broadening of the spectrum,
as well as its shift from the free-atom line center (figure \ref{spectrum3}c).

\begin{figure}[tbp]
\begin{center}
\includegraphics[width=4.05in]{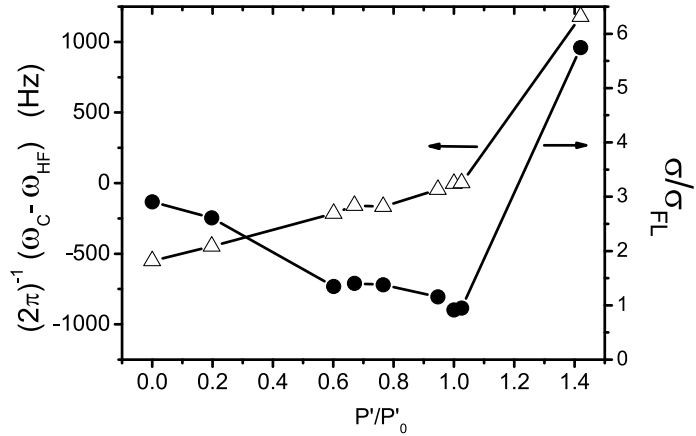}
\end{center}
\caption{Line center ($\omega_{c}$, $\vartriangle$) and RMS width ($\sigma$, $\bullet$) of Rabi spectrum for
trapped atoms as a function of compensating beam power, for a $3 \msec$ pulse. ($\sigma_{FL}\approx110 \hz$ is the
Fourier limited $\sigma$). The spectrum width is minimized to a Fourier-limited value, at a compensating beam
intensity which corresponds also to a minimal shift from $\ohf$. The power is normalized to the measured value at
which the best compensation is achieved, $P'_{0}=25\pm 10 \nw$.} \label{optimum_compensation}
\end{figure}

Figure \ref{optimum_compensation} shows the measured RMS width and shift of the trapped atoms as a function of
compensating beam power, again for a $3 \msec$ microwave pulse. The spectrum width is minimized to the Fourier
broadening limit at a compensating beam power which corresponds also to a minimal shift from the free-atoms line
center.

\begin{figure}[tbp]
\begin{center}
\includegraphics[width=3.55in]{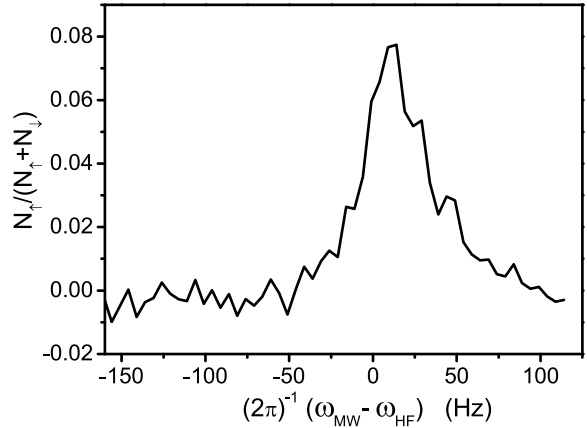}
\end{center} \caption{Rabi spectrum of the hyperfine splitting of
optically trapped $^{85}$Rb, with a $25 \msec$ pulse. A Fourier
limited $\sigma\approx13 \hz$ is measured.} \label{spectrum25}
\end{figure}

Figure \ref{spectrum25} shows the measured spectrum for trapped atoms for a $25 \msec$ long microwave pulse. A
measurement of free atoms with this pulse length is not possible in our setup since the atoms fall due to gravity,
and leave the interaction region. A Fourier limited $\sigma=13 \hz$ is measured, representing a 25-fold reduction
in the line broadening, as compared to the line broadening of trapped atoms. A similar measurement with a $50
\msec$ pulse shows a nearly Fourier limited width (50 times narrower than the trapped atoms spectrum), at the
expense of a much larger spontaneous photon scattering and hence a smaller signal (All four $\ket{F=2, m_F\neq0}$
states are populated and contribute to the spontaneous Raman scattering background, which is hence 5 times larger
than that of an ideal 2-level system). For even longer measurement times spontaneous photon scattering prevents
further narrowing of the line.

We measure the spin relaxation rate \cite{Cline94} to be $\sim 3 \times 10^{-3} \seconds^{-1}$  for atoms in the
trap. The addition of the compensating beam induces an increase of only $\sim 20\%$, as expected.

It should be noted, that for certain transitions a laser frequency can be chosen such as the relative light field
perturbations on the measured spectrum cancel \cite{Katori99, Katori03, Takamoto03}. Although simpler than our
scheme, in the sense that only one laser is needed, ours is a more general method which does not require the
existence of a ``magic wavelength'', where the light shift of the transition vanishes.

\section{Energy Selection}
\label{sec:selection} In this section we present a method for increasing the coherence time in an optical trap, by
narrowing the energy distribution of the trapped ensemble using a microwave ``pre-selection''. First, all the
atoms are optically pumped into $F=3$, where they are distributed among the different $m_F$ states. Then, a weak
``selection'' $\pi$-pulse, with frequency $\omega_{s}$ and duration $t_s$ (and hence FWHM spectral width $\Delta
\omega_{s} / 2 \pi \approx 0.8 / t_s$, see section \ref{sec:classical}), is applied. As a result, a subset of the
atoms in $\ket{\ua}$, composed of those atoms having resonance frequency in the vicinity of $\omega_{s}$, is
transferred to the lower hyperfine state ($\ket{\da}$). Since the resonance frequency is a monotonic function of
energy, this selection can be viewed as an energy selection of atoms in the trap, i.e. an energy band is selected
around the energy $E_{s}$. Next, a strong and short laser pulse resonant with the $\ket{5S_{1/2}, F=3} \rightarrow
\ket{5P_{3/2}, F=4}$ transition ejects all the ``unselected'' atoms from the trap without causing any effect on
the ``selected'' atoms. The end-result of the above sequence is an atomic ensemble with a narrower energy spread,
and hence a longer coherence time.

The number of selected atoms is given by
\begin{eqnarray}
N_s =  \int_0^\infty \, g(E) \, F(E) \, P(\omega_0(E)) \, dE
\end{eqnarray}
where $g(E)$ in the density of states in the trap, $F(E)\propto \exp(-E/k_BT)$ is a Boltzmann factor, and $P$ is
the Rabi transition probability, given in equation \ref{eq:rabi_prob}, and expressed here as a function of
$\omega_0(E)$ to stress the fact that the resonance frequency depends on the energy of the atom in the trap. With
our technique, it is possible to select a narrow energy band around any central energy, enabling, for example, to
maximize the number of selected atoms (for a given energy width) by selecting the energy with the highest density
of thermally populated states. For example, assume a 3D harmonic trap for which the density of states obeys
$g(E)\propto E^{2}$. If the trap is populated with an atomic sample at a temperature $T$, and a selection pulse
corresponding to an energy $E_s$ and (FWHM) width $\Delta E_s \ll k_BT$ is used, then the number of selected atoms
is given by $N_s(E_s)\approx g(E_s) F(E_s) \Delta E_s$. For a given $\Delta E_s$, the selection energy which
optimizes the number of selected atoms is given by $E^{opt}_s=2 k_B T$. A simple selection of the coldest atoms
(e.g. by lowering the trapping potential) will result in a dramatically smaller number of selected atoms. The
ratio between the number of atoms selected around $E^{opt}_s$ and near $E_s=0$ is given by

\begin{eqnarray}
\frac{N(E=E^{opt}_s)}{N(E=0)} \approx \frac{\left(2 k_BT\right)^{2}\eexp{-2}}{\left(\Delta E_s\right)^{2}} \approx
\frac{1}{2} \times \left( \frac{k_B T}{\Delta E_s} \right)^{2}.
\end{eqnarray}

\begin{figure}[tb]
\begin{center}
\includegraphics[width=3.57in] {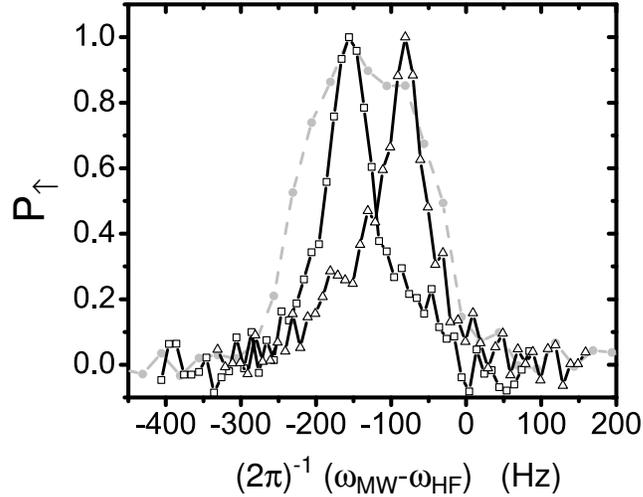}
\caption{Rabi spectrum of energy-selected atoms. Full dots: Spectrum of trapped atoms with no pre-selection. Empty
dots: Rabi spectrum of atoms pre-selected with a $20 \msec$ pulse at a frequency of $\ohf/ 2 \pi -156 \hz$. Empty
triangles: Spectrum of pre-selected atoms with a $20 \msec$ pulse at a frequency of $\ohf/ 2 \pi -81 \hz$. }
\label{fig:selection}
\end{center}
\end{figure}

We perform a proof-of-principle experiment of the energy-selection scheme, using a red-detuned Gaussian trap with
a waist of $w_{0}=50\mcm$, a power of $P=120 \mw$ and a wavelength $\lambda=810 \nm$ (See section \ref{ap:setup}).
In order to probe the resulting energy distribution, we perform Rabi spectroscopy of the remaining atoms using a
$20 \msec$ microwave pulse. As seen in figure \ref{fig:selection}, the spectrum of atoms selected with a short
$100 \mcsec$ pulse (which actually transfers the whole ensemble to $\ket{\da}$) has a width of $\sim 200 \hz$
(FWHM) in agreement with the calculated width for our trap power and detuning and the temperature of the atoms.
The spectrum of atoms selected with a $20 \msec$ pulse has a width of $\sim 60 \hz$ showing a $>3$-fold narrowing
in the energy distribution. Shown in figure \ref{fig:selection} are the results of selection at two different
values of $E_s$, demonstrating the ability to maximize the number of selected atoms, as explained above.

In a similar way to the ``echo'' spectroscopy of section \ref{sec:echo} below, any broadening of the narrow
selected energy-slice as a consequence of different effects, such as photon scattering and trap instabilities, can
be very instructive on the effects causing the broadening. For example, a change in the trapping potential after
the selection pulse and before the measurement, will result in a line-shape that reflects the trap's local density
of states (LDOS) \cite{Cohen01}.

\section{Microwave Spectroscopy in a Dipole Trap: A Quantum Model}
\label{sec:quantum}
\begin{figure}[tb]
\begin{center}
\includegraphics[width=3.1in] {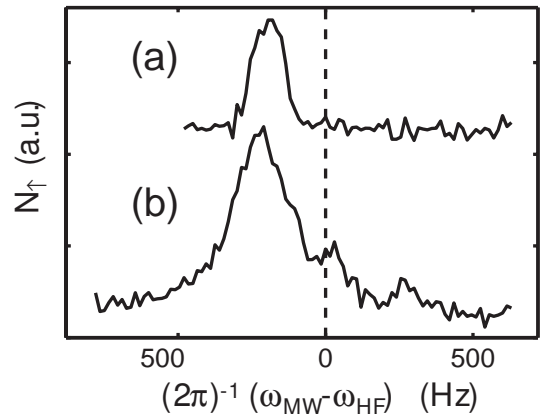}
\caption{Rabi spectroscopy of atoms in a trap with wavelength $\lambda=805 \nm$, with a 20 ms pulse. (a) For a
microwave pulse area corresponding to a $ \pi $-pulse the sidebands are not visible, indicating that the matrix
elements for the sidebands are very small. (b) For a stronger pulse, the sidebands emerge.} \label{echo_sidebands}
\end{center}
\end{figure}

The classical model of section \ref{sec:classical} fails to completely describe the spectroscopic properties of
the trapped atoms. As an example, figure \ref{echo_sidebands} shows that for a powerful and long enough pulse, the
resonance line in a Rabi spectrum of trapped atoms develops sidebands, which are, as we shall show next, a clear
evidence of the quantization of the motional levels in the trap.

In reference \cite{Davidson95}, a semi-classical, dynamic model was used, in which the center-of-mass motion of
the atoms is solved, and the phase is integrated over the classical trajectory of the atoms. The total signal is
obtained by averaging over the different trajectories. The validity of this model is not clear, since it assumes
the same classical trajectory for particles in different internal states. Such a model ignores the separation of
trajectories with same initial conditions due to the difference in potentials, and predicts longer coherence times
than the measured ones \cite{Davidson95}.

\begin{figure}[tb]
\begin{center}
\includegraphics[width=2in] {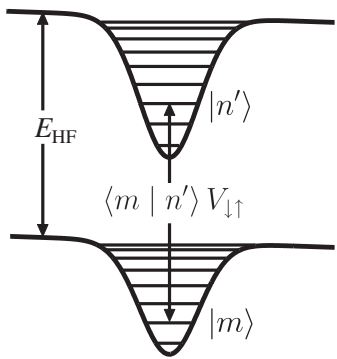}
\caption{The eigenenergies of the trapped atoms consist of two manifolds (belonging to $\left| \downarrow
\right\rangle $ and $\left| \uparrow \right\rangle $) separated in energy by $\ehf$. The matrix elements for
microwave transitions depend on the overlap of the vibrational states. The trap shown is a Gaussian trap, and
gravity gives a small slope.} \label{trap_levels}
\end{center}
\end{figure}

We present in this section a quantum-mechanical treatment of microwave spectroscopy in an optical dipole trap. The
existence of a difference in potential for atoms in different hyperfine states, means that the external (center of
mass) potential depends on the internal (spin) state, hence the internal and external degrees of freedom cannot be
separated and the entire Hamiltonian including both internal and external degrees of freedom has to be considered:
\begin{eqnarray}
\mathcal{H}&=&H_{\downarrow }\left| \downarrow \right\rangle \left\langle \downarrow \right| +\left( H_{\uparrow
}+\ehf\right) \left| \uparrow \right\rangle \left\langle \uparrow \right| \nonumber \\
&=&\left[ \frac{p^{2}}{2m}+V_{\downarrow }\left( {\bf x}\right) \right] \left| \downarrow \right\rangle
\left\langle \downarrow \right| +\left[ \frac{p^{2}}{2m}+V_{\uparrow }\left( {\bf x} \right) +\ehf\right] \left|
\uparrow \right\rangle \left\langle \uparrow \right| \label{ham},
\end{eqnarray}
where $ V_{\downarrow }$ and $V_{\uparrow }$ are the external potentials for atoms in states $\left| \downarrow
\right\rangle $ and $\left| \uparrow \right\rangle $, respectively. These potentials include the gravitational
potential, equal for both states, and the dipole potential, which can be written as $U_{\downarrow }$ and
$U_{\uparrow }=(1+\epsilon)U_{\downarrow }$, where $\epsilon\equiv\omega_{HF}/\delta$ can be called the
``perturbation strength'', typically $10^{-3}-10^{-2}$ in our experiments. We assume that the potential can be
approximated as an harmonic one. Since $\ehf$ is much larger than both $ V_{\downarrow }$ and $V_{\uparrow }$, the
eigenenergies of the above Hamiltonian consist of two manifolds separated by $E_{HF}$, and composed of
``vibrational'' levels with a separation of $\hbar\omega _{osc}$ (where $\omega _{osc}$ is the oscillation
frequency in the trap) \footnote{We use here one-dimensional notations, since the dynamics in our trap is
separable. In section \ref{sec:chaos} and references \cite{Andersen04b, Andersen04c} we treat non-separable
(chaotic) dynamics.}. We enumerate the eigenstates of the lower manifold using regular numbers and those from the
upper manifold using primed numbers (see figure \ref{trap_levels}).

When a microwave field $\vmw$ with frequency close to $\ohf$ is applied, transitions between the eigenstates of
the Hamiltonian corresponding to different internal states, are driven\footnote{Since the size of our trap ($\sim
50\mcm$) is much smaller than the microwave wavelength ($\sim 10\cm$), the momentum of the microwave photon can be
neglected (Lamb-Dicke regime \cite{Cohen-Tannoudji92})}. The microwave field acts only on the internal state of
the atoms, hence the matrix elements for the transitions can be written as the free-space matrix elements for the
internal state transition, times the overlap between the initial and final vibrational eigenstates, which
correspond to different Hamiltonians:
\begin{eqnarray}
\braOket{m,\da}{\vmw}{n',\ua}&=&\braket{m}{n'} \cdot \braOket{\da}{\vmw}{\ua}\equiv\braket{m}{n'} V_{\da \ua}\nonumber\\
\braOket{m',\ua}{\vmw}{n,\da}&=&\braket{m'}{n} \cdot \braOket{\ua}{\vmw}{\da}\equiv\braket{m'}{n} V_{\ua \da}\nonumber\\
\braOket{m',\ua}{\vmw}{n',\ua}&=&\braket{m'}{n'} \cdot \braOket{\ua}{\vmw}{\ua}=0\nonumber\\
\braOket{m,\da}{\vmw}{n,\da}&=&\braket{m}{n} \cdot \braOket{\da}{\vmw}{\da}=0,\label{mat_el}
\end{eqnarray}
where we have used the orthogonality of the$\{\ket{n}\}$ and
$\{\ket{n'}\}$ states.

Now we can understand the result of figure \ref{echo_sidebands}. Non-vanishing matrix elements for transitions to
vibrational states with $n^{\prime}\neq n$ show up as sidebands in the microwave spectrum, if the pulse is long
and weak enough so that the power broadening is smaller than the typical level spacing. For a microwave pulse area
corresponding to a $ \pi $-pulse (figure \ref{echo_sidebands}a) the sidebands are not visible, indicating that the
matrix elements for the sidebands are very small. Only for a stronger pulse (figure \ref{echo_sidebands}b), the
sidebands emerge. We verified that the sidebands are stronger for smaller values of the detuning, and hence larger
perturbation and larger $\braket{m}{n'}$ for $m \neq n'$, as expected from the analysis. The coupling to different
vibrational levels, has a strong analogy to the Franck-Condon factors in molecular spectroscopy. Note, that the
near harmonicity of our trap is responsible for the fact that the sidebands are seen even though many levels are
thermally populated.

We start by considering a quantum state written as the external product of an internal state and a state which
represents the ``external'' degree of freedom. If the atoms are initially prepared in their internal ground state
$\ket{\da}$, their total wavefunction can be written as $\Psi =\left| \downarrow \right\rangle \otimes \psi$ or,
shortly, $\ket{\da,\psi}$. In the presence of the microwave field, we need to solve the time-dependent
Schr\"{o}dinger equation given by
\begin{eqnarray}
\left[\mathcal{H}+ \vmw\right] \Psi = i \frac{\partial \Psi}{\partial t} \label{se},
\end{eqnarray}
with the Hamiltonian of equation \ref{ham}. We choose to use $\hbar \equiv 1$ to simplify the equations. As is
usually done in time-dependent perturbation theory, we choose to work in the basis of eigenstates of the
Hamiltonian without the microwave field (equation \ref{ham}). A general state can be written as
\begin{eqnarray}
\Psi=\sum_n a_n \ket{n,\da}+\sum_{n'} a_{n'} \ket{n',\ua}.
\end{eqnarray}
Introducing this into equation \ref{se}, results in
\begin{eqnarray}
\sum_n a_n (E_n+\vmw) \ket{n,\da} + \sum_{n'} a_{n'} (E_{n'}+\ehf+\vmw) \ket{n',\ua}\nonumber\\
=i \left[ \sum_n \dot{a}_n \ket{n,\da} + \sum_{n'} \dot{a}_{n'} \ket{n',\ua} \right].
\end{eqnarray}
Two equations are obtained by projecting the above equation onto $\bra{m,\da}$ and $\bra{m',\ua}$:
\begin{eqnarray}
i  \dot{a}_m&=&a_m E_m+\sum_n a_n \braOket{m,\da}{\vmw}{n,\da} + \sum_{n'} a_{n'}
\braOket{m,\da}{\vmw}{n',\ua}\\
i  \dot{a}_{m'}&=&a_{m'} (E_{m'}+\ehf) + \sum_{n'} a_{n'} \braOket{m',\ua}{\vmw}{n',\ua} + \sum_{n} a_{n}
\braOket{m',\ua}{\vmw}{n,\da}\nonumber.
\end{eqnarray}
Using the matrix elements from equation \ref{mat_el}, these equations can be written as
\begin{eqnarray}
i  \dot{a}_m&=&a_m E_m+ \sum_{n'} a_{n'} \braket{m}{n'} V_{\da \ua}\nonumber\\
i  \dot{a}_{m'}&=&a_{m'} (E_{m'}+\ehf)+\sum_n a_n \braket{m'}{n} V_{\ua \da}.
\end{eqnarray}
We assume now a monochromatic, linearly polarized, microwave field. Then we can write
\begin{eqnarray}
V_{\ua \da}=-\half  \left[\Omw \exp(-i \omw t) + \Omw \exp(i \omw t)\right]\label{field},
\end{eqnarray}
where $\Omw$ is the microwave field Rabi frequency. It is now useful to define a new amplitude, which ``rotates''
at the field's frequency, $b_{n'}=a_{n'} \exp(i \omw t)$. In this new frame the equations are
\begin{eqnarray}
i  \dot{a}_m&=&a_m E_m+ \sum_{n'} b_{n'} \braket{m}{n'} V_{\da \ua} \exp(-i \omw t)\nonumber\\
i  \dot{b}_{m'}&=&b_{m'} (E_{m'}+\ehf- \omw)+\sum_n a_n \braket{m'}{n} V_{\ua \da} \exp(i \omw t)\label{rot}.
\end{eqnarray}
When equation \ref{field} is introduced into equations \ref{rot}, the resulting expression includes terms with a
$\exp(-i 2 \omw t)$ time-dependence. These terms oscillate so fast, compared to every other time variation in the
equations that they can be assumed to average to zero over any realistic time interval, and can be neglected
(rotating-wave approximation). This results in
\begin{eqnarray}
i  \dot{a}_m&=&a_m E_m - \half  \Omw \sum_{n'} \braket{m}{n'} b_{n'}\nonumber\\
i  \dot{b}_{m'}&=&b_{m'} (E_{m'}+  \Dmw) - \half  \Omw \sum_n \braket{m'}{n} a_n,\label{comp}
\end{eqnarray}
where $ \Dmw\equiv \ehf- \omw$ is the microwave detuning.

\section{Ramsey Spectroscopy in a Dipole Trap}
\label{sec:ramsey}

In general, each of the eigenstates in the $\ket{\downarrow}$-manifold is coupled to many eigenstates in the
$\ket{\uparrow}$-manifold by the microwave field, hence there is no simple prediction of the wave function after
microwave irradiation. However, in order to calculate the outcome of a Ramsey experiment (consisting of two
microwave pulses of short width $t$, separated by a much longer ``dark'' period of duration $\tau$), we only need
to solve equations \ref{comp} in two limiting cases: first, the short and strong limit (i.e. $t \ll
\omega_{osc}^{-1}$ and $ \Omw \gg \Dmw + (E_{n}-E_{n'})$ for all $n,n'$), and next, the free evolution limit
(where $\Omw=0$).

In the simple limit of short microwave pulses equations \ref{comp} reduce to
\begin{eqnarray}
i \dot{a}_m&= - \half \Omw^* \sum_{n'} \braket{m}{n'}  b_{n'}\\
i \dot{b}_{m'}&= - \half \Omw \sum_n \braket{m'}{n} a_n.
\end{eqnarray}
Both $\left\{\ket{n}\right\}$ and $\left\{ \ket{n'}\right\}$ are complete basis that span the vibrational part of
the wave function. Hence, we can express the vibrational part of $\ket{n',\ua}$ using $\left\{ \ket{n}\right\}$.
The coefficients in these expressions, denoted $c_n$, are given by $c_n=\sum_{n'} \braket{n}{n'}  b_{n'}$. Using
also the fact that $b_{n'}=\sum_{n} \braket{n'}{n}  c_{n}$, and some algebra, it can be shown that
\begin{eqnarray}
i \dot{a}_m&= - \half \Omw^* c_{m}\\
i \dot{c}_{m}&= - \half \Omw a_m.
\end{eqnarray}
These last equations show that there is no coupling between $a_m$ and $c_n$ for $m \neq n$, hence the solution of
the coupled equations for every $m$ is given by equation \ref{eq:rabi_prob}. This means that if we start with a
state $\ket{\da ,\psi} = \ket{\psi} \otimes \ket{\da}$, the state after an on-resonance pulse of duration $t$ is
given by $\ket{\psi} \otimes [\cos \left(\Omw t/2 \right) \ket{\da} + i \sin \left(\Omw t/2 \right) \ket{\ua}]$.
Hence, for a $\pi$ pulse the vibrational part of the wavefunction is only ``projected'' on the other potential and
will result in $i \ket{\uparrow ,\psi} $. A $\frac{\pi }{2}$-pulse will result in the coherent superposition state
$\halfroot \left( \ket{\downarrow ,\psi}+ i \ket{\uparrow ,\psi} \right)$.

For the ``dark'' periods we go back to equation \ref{comp}, using $\Omw=0$, hence
\begin{eqnarray}
i  \dot{a}_m&=&a_m E_m\\
i  \dot{b}_{m'}&=&b_{m'} (E_{m'}+  \Dmw),
\end{eqnarray}
which means that the time evolution of $\ket{\psi,\ua}$, and $\ket{\psi,\da}$ is given by
\begin{eqnarray}
\ket{\psi(t),\ua}&=& \exp\left[- i (H_{\ua}+ \Dmw)\,t\right] \ket{\psi(0),\ua}\\
\ket{\psi(t),\da}&=& \exp\left[- i H_{\da}\,t\right] \ket{\psi(0),\da}.
\end{eqnarray}

We are now prepared to calculate the result of a Ramsey experiment. Let us assume that atoms are initially
prepared in their internal ground state, and also that their vibrational part is an eigenstate of $V_{\downarrow}$
characterized by the quantum number $n$. Their wavefunction can then be written as $\ket{\psi_n,\downarrow}$,
where we use the notation $\psi_n$ instead of $\ket{n}$ to stress that $\ket{\psi_n,\ua}$ is not equal to
$\ket{n,\ua}$. The atom is irradiated with a short $\pit$-pulse to generate the wave function $\halfroot \left(
\ket{\psi_n,\downarrow}+ i \ket{\psi_n, \uparrow} \right)$. After some time $\tau $ this state will, in the
rotating frame of the microwave field, evolve into
\begin{eqnarray}
\frac{1}{\sqrt{2}} \exp\left[-iH_{\da}\tau\right] \ket{\psi_n,\downarrow} + \frac{i}{\sqrt{2}} \exp\left[-i \left(
H_{\ua}+\Dmw \right) \tau\right] \ket{\psi_n, \uparrow}.
\end{eqnarray}
Then the atoms are irradiated with a second $\frac{\pi }{2}$-pulse generating the wave function
\begin{eqnarray}
\lefteqn{\frac{1}{2} \left\{ \exp\left[-iH_{\da}\tau\right] - \exp\left[-i\left(
H_{\ua}+\Dmw \right) \tau\right] \right\} \ket{\psi_n,\downarrow}} \nonumber\\
& & + \frac{i}{2} \left\{ \exp\left[-i\left( H_{\ua}+\Dmw \right) \tau\right] + \exp\left[-iH_{\da}\tau\right]
\right\} \ket{\psi_n, \uparrow}.
\end{eqnarray}
The population of state $\ket{\uparrow}$ is then,
\begin{eqnarray}
P_{\ua}&=&\frac{1}{4}\left\{\bra{\psi_n} \left[ \eexp{i\left( H_{\ua}+\Dmw \right) \tau } + \eexp{i H_{\da} \tau}
\right] \left[ \eexp{-i\left( H_{\ua}+\Dmw \right) \tau} + \eexp{-i H_{\da} \tau} \right] \ket{\psi_n} \right\},
\end{eqnarray}
which can also be written as
\begin{eqnarray}
P_{\ua}=\half \left\{ 1+ \textrm{Re} \left[ \braOket{\psi_n}{\eexp{i \left(H_{\ua} + \Dmw \right) \tau} \eexp{-i
H_{\da} \tau}}{\psi_n} \right] \right\}.
\end{eqnarray}
The expression inside the square brackets is, in general, a complex number which can be written as $A\exp (i \Dmw
\tau + \phi_n)$, where $\phi_n(\tau)$ is a slowly varying phase depending on the initial state $\left|
\psi_n\right\rangle $ and the dynamics of it in the trap. Then we can write
\begin{eqnarray}
P_{\ua}=\half \left( 1+ \left| \braOket{\psi_n}{\eexp{i H_{\da} \tau} \eexp{-i H_{\ua} \tau}}{\psi_n} \right| \cos
\left[ \Dmw \tau + \phi_{n} \left( \tau \right) \right] \right). \label{ram}
\end{eqnarray}

Equation \ref{ram} shows that, when $\Dmw$ is sufficiently large compared to variations in $\phi_{n} \left( \tau
\right)$, scanning $\omw$ for a fixed $\tau $ yields the usual Ramsey fringes with a contrast given by $\left|
\braOket{\psi_n}{\eexp{i H_{\da} \tau} \eexp{-i H_{\ua} \tau}}{\psi_n} \right|$. The fringes contrast is actually
the ``fidelity'' of the external motion in the trap, since it can be viewed as the overlap between a ``desired''
state $\eexp{-i H_{\da} \tau}\ket{\psi_n}$ and the actual state in a perturbed environment $\eexp{-i H_{\ua}
\tau}\ket{\psi_n}$. The fidelity was first proposed by Peres \cite{Peres84} as an indicator of the stability of a
quantum system, in an analogue way to characterizing the stability of a classical system with the Lyapunov
exponent \cite{Schuster84}. Alternatively, the Ramsey fringes contrast can be interpreted as a Loschmidt echo
\cite{Jalabert01}, that measures the overlap of a state evolved forward in time (with $H_{\da}$), and then
backward in time with a perturbed hamiltonian $H_{\ua}$. Finally, since the initial wavefunction is an eigenstate
of $H_{\da}$, the contrast of the Ramsey fringes can also be written as a time-correlation function
$\left|\braket{\psi\left(t=0\right)}{\psi\left(t=\tau \right)}\right|$. If $ V_{\da}\left( {\bf x}\right)
=V_{\ua}\left( {\bf x}\right) $ then an eigenstate of $H_{\da}$ is also an eigenstate of $H_{\ua}$, and clearly
this contrast equals unity. In general, if we start with an eigenstate of $H_{\da}$, the projected state will not
be an eigenstate of $H_{\ua}$, and therefore will evolve in the new potential. Such quantum dynamics causes the
overlap $\braket{\psi\left(t=0\right)}{\psi\left(t=\tau \right)}$ to decay in an interesting way, which depends on
the type of the underlying classical dynamics (being regular, chaotic or mixed) and the strength and type of the
perturbation \cite{Andersen04b}.

Such quantum dynamics and in particular the decay of fidelity or Loschmidt echo in chaotic systems has been the
topic of intense theoretical and numerical studies in recent years (see for example references
\cite{Cucchietti00,Jacquod01, Cerruti02,Cucchietti02b,Cucchietti02,Prosen02, Emerson02, Vanicek03,Wisniacki03,
Benenti03, Hiller04}), mainly because fidelity is also the standard measure for loss of information in quantum
computation \cite{Nielsen01}. However, experimental studies of chaotic and mixed systems are still a missing
chapter, since they require the preparation of highly-excited, pure quantum states to avoid ``spreading'' these
interesting effects by averaging over an inhomogeneously broadened system.

This  difficulty is clearly manifested in our experimental system, composed of a thermal ensemble of atoms
incoherently populating more than $10^{6}$ eigenstates (as opposed to an initial single vibrational eigenstate,
considered so far). The total population in $\ket{\ua}$ is now given by an average of $P_{\ua}$ over the initial
thermal ensemble. If we assume that $\left| \braOket{\psi_n}{\eexp{i H_{\da} \tau} \eexp{-i H_{\ua} \tau}}{\psi_n}
\right|\equiv C$ does not depend on $n$, then the total population is given by
\begin{eqnarray}
P_{\ua}=\half \left( 1+  C \sum_n F(n) \cos \left[ \Dmw \tau + \phi_{n} \left( \tau \right) \right] \right).
\end{eqnarray}
where $F(n)=\exp(-E_n/ k_B T) / \sum_n \exp(-E_n/ k_B T)$ is a Boltzmann factor.

We have shown in reference \cite{Andersen04b} that in the small-perturbation regime, achieved experimentally for
large values of the detuning, an eigenstate of $H_{\da}$ (in the lower manifold) is coupled by the microwave field
mostly to the corresponding eigenstate of $H_{\ua}$ (in the upper manifold), and hence the fidelity of each state
is nearly unity. However, since $\phi_{n} \left( \tau \right)=(E_n-E_{n'=n})t$ depends on the initial state, the
ensemble-averaged fringe contrast in a Ramsey experiment will decay rapidly. For this small-perturbation regime,
the system acts as an inhomogeneously broadened ensemble of noninteracting two-level systems and thus the
Ramsey-fringe decay time can be simply estimated by $1/(2\Delta _{RMS}$), where $\Delta _{RMS}$ is the RMS spread
of the resonance frequencies for $\left| n\right\rangle \rightarrow \left| n'=n\right\rangle $ transitions, taken
over the thermal ensemble.

\begin{figure}[tbp]
\begin{center}
\includegraphics[width=3.65in]{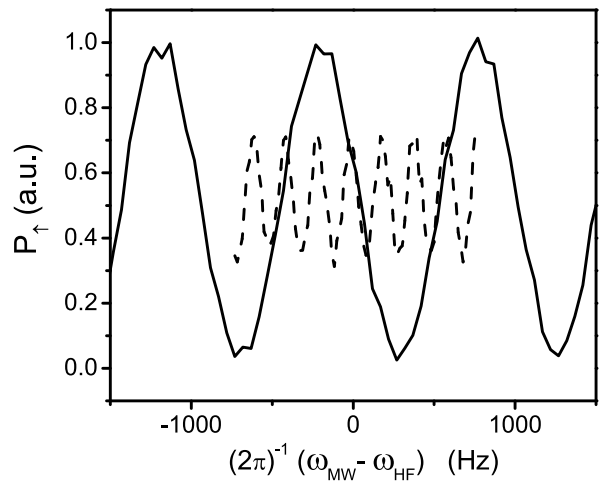}
\end{center} \caption{Ramsey spectrum of trapped atoms, measured for $\tau=1 \msec$ (full line) and $\tau=5 \msec$ (dashed line).
The ensemble-averaged fringe contrast decays rapidly because the cosine terms from different populated states get
out of phase.} \label{fig:ramsey}
\end{figure}

As an example, figure \ref{fig:ramsey} shows Ramsey fringes of trapped atoms, as measured in our experimental
setup. The atoms are loaded into a $50 \mw$ horizontal laser beam focused to a 1/e$^{2}$ radius of $50 \mcm$, and
with a wavelength of $ \lambda=800 \nm$, yielding a trap depth of $U_{0}/k_{b}T=1.5$ (See
 section \ref{ap:setup}). Under these conditions the fidelity of each eigenstate as nearly unity. Nevertheless, the
contrast of the Ramsey fringes at $\tau=5 \msec$ is reduced to $\sim 1/3$ of its value at $\tau=1 \msec$. In Fig
\ref{echo_vs_ramsey}, this measured Ramsey fringe contrast is shown as a function of time. As seen, the Ramsey
fringe contrast decays on a time scale of $2.4 \msec$, due to the variation of $\phi_{n}$ over the thermally
populated states. This results are in agreement with a calculated decay time of 2.7 ms, assuming a thermal
ensemble in a harmonic trap clipped at 1.5 $k_{B}T$.

In conclusion to this section, Ramsey spectroscopy of optically trapped atoms occupying a pure state can yield
important information on their quantum dynamics and in particular on the fidelity (or equivalently the Loschmidt
echo) as they are perturbed by the small and well controlled difference in optical potential acting on the two
internal states. However, in most experiments and in ours in particular, pure and highly excited states are
practically inaccessible, and for thermal ensembles the interesting quantum effects are overwhelmed and completely
smeared out by the rapid inhomogeneous dephasing of the system. In the next section we show how an echo-like
scheme can suppress this inhomogeneous dephasing and enable us to directly measure the quantum dynamics of thermal
ensembles of optically trapped atoms.

\section{Echo Spectroscopy}
\label{sec:echo}

\begin{figure}[t]
\begin{center}
\includegraphics[width=3in]{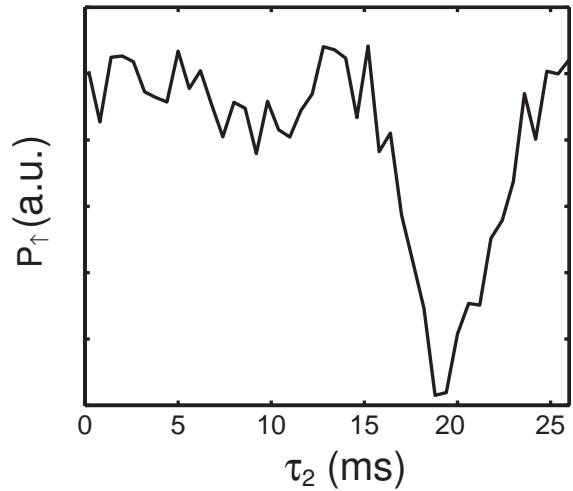}
\end{center}
\caption{Echo signal ($P_{\ua}$) measured as a function of the time $\protect\tau_{2}$ between the $\protect\pi
$-pulse and the second $\pit$-pulse, for a fixed $\protect \tau_{1}$. A dip in $P_{\ua}$ is seen, showing a
``coherence echo'' at $\protect\tau_{2}=\protect\tau_{1}$.} \label{fi1}
\end{figure}

The results of the previous section indicate that the  decay of the Ramsey fringe contrast is not only a
fingerprint of decoherence (an irreversible effect) but also a consequence of dephasing. Dephasing, which causes
the ensemble averaged signal to decay, can be reversed, at least partially, by stimulating an effective ``time
reversal'', as has been reported for spin echoes \cite{Hahn50} and photon echoes \cite{Kurnit64,Allen87}, and more
recently for a motional wave packet echo using ultra cold atoms in a one-dimensional optical lattice
\cite{Buchkremer00}. We achieve such reversal in dephasing by adding a $\pi $-pulse, which inverts the populations
of $\ket{\da}$ and $\ket{\ua}$, between the two $\frac{\pi }{2}$ pulses \cite{Andersen03}. If the $\pi $-pulse is
exactly in the middle between the two $\frac{\pi }{2}$-pulses the two parts of the superposition state generated
by the first $\frac{\pi }{2}$-pulse spend an equal amount of time in both levels, and are therefore exactly in
phase with each other at the time of the second $ \frac{\pi }{2}$-pulse. This means that a ``coherence echo''
appears at the time of the second $\frac{\pi }{2}$-pulse even for a system that has dephased completely before the
$\pi $-pulse. The coherence echo is observed by seeing that all the atoms return to the initial state after the
$\frac{\pi }{2}$-$\pi $-$\frac{\pi }{ 2}$-pulse sequence. If the time $\tau _{1}$ between the first $\frac{\pi }{
2}$ and the $\pi $-pulse is kept constant, and the time $\tau _{2}$ between the $\pi $ and the second $\frac{\pi
}{2}$-pulse is swept, the coherence echo is seen as a decrease in the population of $\ket{\ua}$ for $\tau
_{2}=\tau _{1}$ (see figure \ref{fi1}). The population $P_{\ua}$ for $\tau _{2}=\tau _{1}$ is referred in what
follows as the ``echo signal''\, where $P_{\ua}=0$ indicates full revival of coherence and $P_{\ua}=\frac{1}{2}$
indicates complete dephasing.

A similar calculation as the one in the previous section shows that the echo signal for an initial state
$\ket{\psi_n}$ is now
\begin{eqnarray}
P_{\uparrow} &=&\frac{1}{2}\left[ 1- \mathop{\rm Re} \left( \braOket{\psi_n}{\, \eexp{i H_{\da} \tau} \,\eexp{i
H_{\ua} \tau} \,\eexp{-i H_{\da} \tau} \,\eexp{-i H_{\ua} \tau}\,}{\psi_n} \right) \right]. \label{formel1}
\end{eqnarray}
The main difference between the echo signal (equation \ref{formel1}) and the Ramsey signal (equation \ref{ram}) is
that for the echo $P_{\uparrow}$ no longer depends on $\Dmw$. Hence, when measuring a thermal ensemble, the
condition $\left\langle n'\mid n\right\rangle \simeq \delta_{nn'}$ for all initially populated vibrational states
now ensures that $P_{\ua}\simeq 0$ for all $\tau $ \cite{Andersen04b}. In other words, after dephasing for a time
$\tau $ a $\pi $-pulse stimulates a {\it coherence echo} of the ensemble averaged signal at time 2$\tau $.

Some intuition can be gained, by switching to a new basis $\{\varphi_k\}$, defined by $\ket{\varphi_k(t=0)} =
\exp(-i H_{\ua} \tau)\ket{\psi_n}$. Now, the echo signal can be written using a time correlation function:
\begin{eqnarray}
P_{\uparrow} &=&\frac{1}{2}\left\{ 1- \mathop{\rm Re}\left[ \exp\left(i E_n \tau\right)
\braket{\varphi_k(t=0)}{\varphi_k(t=\tau)}\right] \right\},\label{tcf}
\end{eqnarray}
where $\ket{\varphi_k(t=\tau)} = \exp(-iH_{\da} \tau)\ket{\varphi_k(t=0)}$.

\begin{figure}[tb]
\begin{center}
\includegraphics[width=3.6in] {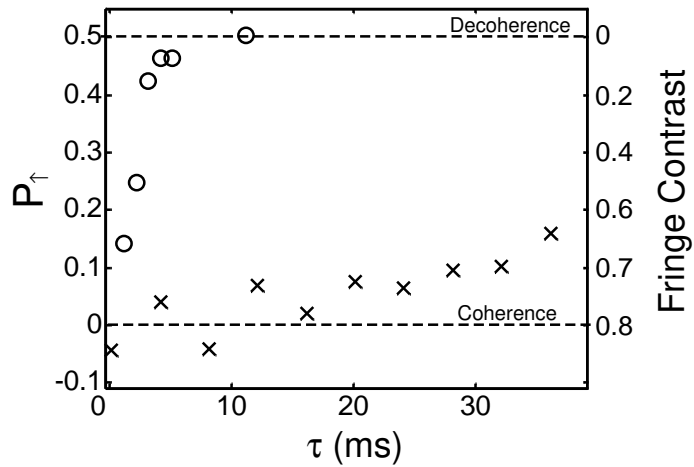}
\caption{Ramsey fringe contrast ($\circ$) and echo signal $P_{\ua}$ ($\times$) measured as a function of the time
between the two $\pit$-pulses. For the echo signal the value 0 represents complete coherence, and the value 1/2
represents complete dephasing. The trap laser wavelength is 800 nm. A coherent echo ($P_{\ua}<<1/2$) persists long
after the Ramsey fringe contrast has decayed.} \label{echo_vs_ramsey}
\end{center}
\end{figure}

The results of echo spectroscopy with the same trap parameters as used before for Ramsey spectroscopy in section
\ref{sec:ramsey} are also presented in figure \ref{echo_vs_ramsey}. For these parameters $\left\langle n'\mid
n\right\rangle \simeq \delta_{nn'}$ is a good approximation for all thermally populated states. As explained
before, we subtract from the signal contributions to the population of $\ket{\ua}$ due to $F$-changing Raman
transitions induced by the trap laser and normalize to the signal after a short $ \pi $-pulse, which transfers the
whole population of $\ket{\da}$ to $\ket{\ua}$. This corrected signal is denoted $P_{\ua}$. A coherence echo
($P_{\ua}<<1/2$) is clearly seen long after the Ramsey fringe contrast has decayed. On a time scale of $\sim $100
ms the echo coherence decays, for reasons that will be analyzed further in section \ref{ssec:manypi}.

\section{Quantum Dynamics} \label{ssec:qdynamic}
Since echo spectroscopy is seen to cope with the inhomogeneous broadening, we can use it to observe the
fingerprints of the fidelity even for a thermal ensemble. When $\left\langle n'\mid n\right\rangle \neq \delta
_{nn'} $ a good echo signal is no longer expected, since each vibrational state is coupled to several vibrational
states by the microwave fields, and therefore $ \left| \left\langle \varphi _{n}\left( t=0\right) \mid \varphi
_{n}\left( t=\tau \right) \right\rangle \right| <1$.

\begin{figure}[tb]
\begin{center}
\includegraphics[width=3.1in] {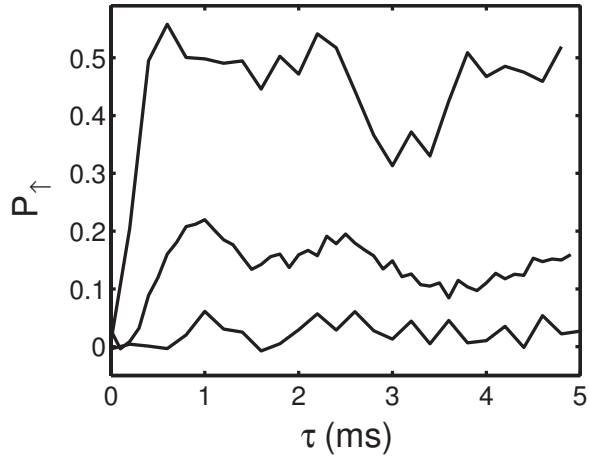}
\caption{Echo signal $P_{\ua}$ measured for three different trap laser wavelengths $\protect\lambda $. Lower
curve: $\protect\lambda=805 \nm$. A good echo signal ($P_{\ua}<<1/2$) is seen almost independent of time between
pulses. Middle curve: $\protect\lambda=798.25 \nm$. The echo signal oscillates around a value smaller than 1/2
with partial revivals at $ \tau=\tau_{osc}/2$ and $\tau=\tau_{osc}$, where $\tau_{osc}=3.6 \msec$ is the measured
trap oscillation frequency in the transverse direction. Upper curve: $\protect \lambda=796.25 \nm$. After a short
time the echo signal completely disappears ($P_{\ua}=1/2$), but partly revives again at $\tau=3.3 \msec$, close to
$\tau_{osc}$.} \label{echo_revivals}
\end{center}
\end{figure}

Ignoring gravity, $V_{\da}$ and $V_{\ua}$ are just the dipole potentials, and are related by $V_{\ua} =
(1+\varepsilon) V_{\da}$, where $\varepsilon=\ohf/\delta$. Hence, the detuning serves as a control parameter for
the perturbation strength. We perform echo spectroscopy as a function of time between pulses for different
wavelengths of the trap laser (and hence different perturbation strengths, $\varepsilon$) while keeping the trap
depth constant by adjusting the power of the trap beam. The results are shown in figure \ref{echo_revivals}. For a
large detuning ($\lambda=805 \nm$, which means a ``weighted detuning'' $\delta=16.7 \nm$ and hence a perturbation
strength $\varepsilon\sim 4\times 10^{-4}$) a good echo ($P_{\ua}<<1/2$) is seen independent of the time between
pulses, as also seen in figure \ref{echo_vs_ramsey}. For an intermediate detuning ($\lambda=798.25 \nm$,
$\varepsilon=9\times10^{-4}$) damped oscillations to a level smaller than 1/2 are seen, and for a small detuning
($\lambda=796.25 \nm$, $\varepsilon=2\times10^{-3}$) a complete decay of the echo coherence ($P_{\ua}=1/2$) is
observed, followed by partial revivals at later times.

\begin{figure}[tb]
\begin{center}
\includegraphics[width=3.1in] {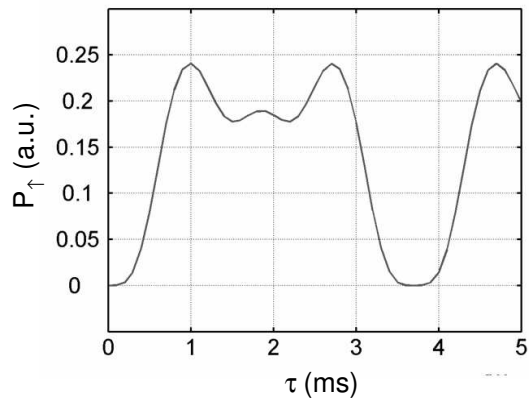}
\caption{Numerical calculations of the echo signal in a 2D harmonic oscillator, with gravity, having the same
oscillation frequency as the experimental data in figure \ref{echo_revivals}. A similar calculation shows that
without the effect of gravity, the revival at $\tau =1.8 \msec$ is complete.} \label{echo_sim}
\end{center}
\end{figure}
\begin{figure}[tb]
\begin{center}
\includegraphics[width=3.3in] {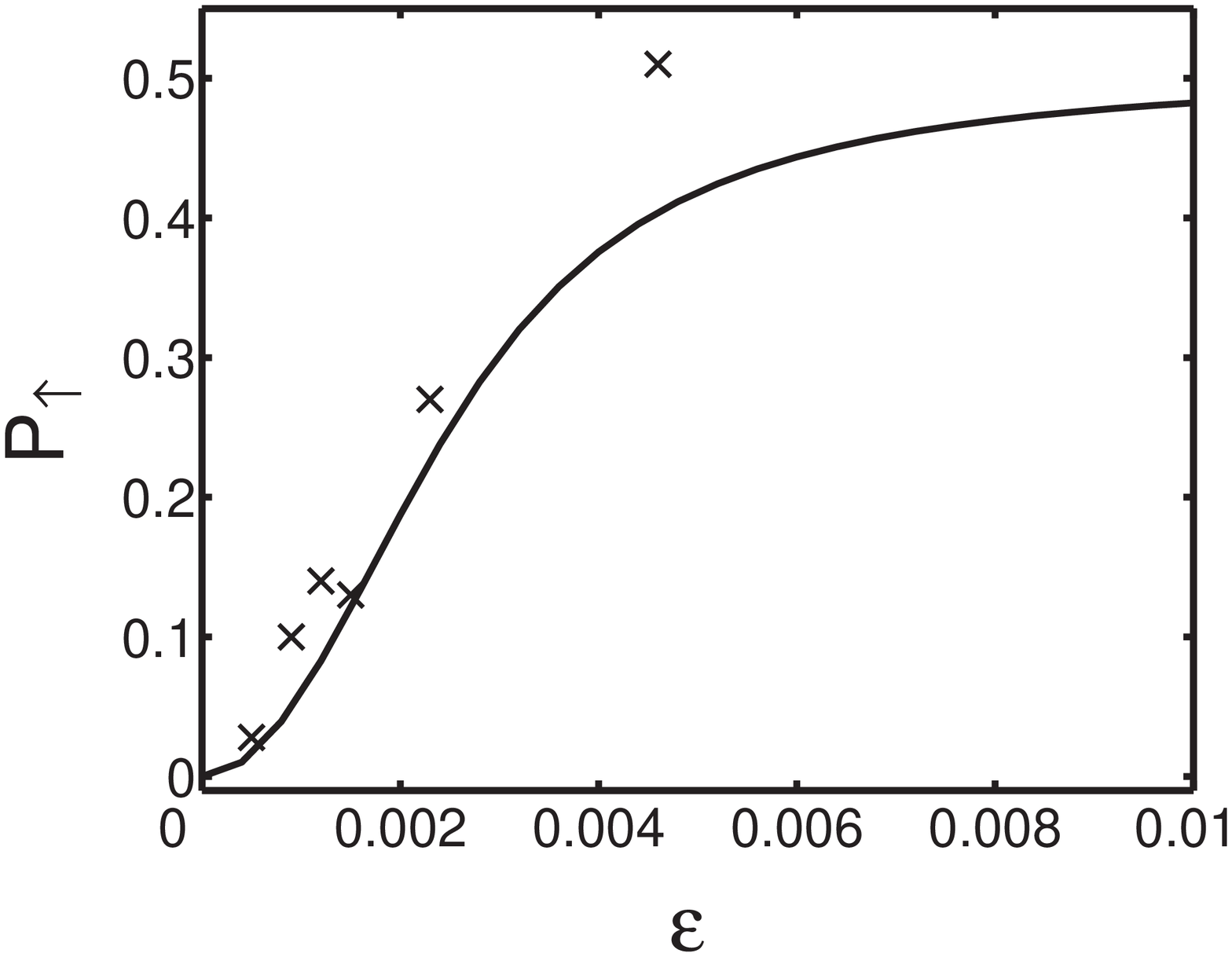}
\caption{Long time level of echo signal as a function of $\varepsilon$, the relative difference between
potentials. ($\times$): Experimental measurement. Solid line: Calculation of the ensemble average of $\left|
\left\langle n^{\prime}=n \mid n\right\rangle \right| ^{4}$ for a 2D harmonic potential with oscillation time of
$3.6 \msec$.} \label{echo_long_time}
\end{center}
\end{figure}

The interpretation for the large detuning (small perturbation) regime, where $\left\langle n'\mid n\right\rangle
\simeq \delta _{nn'} $, was given above. For an intermediate detuning, a small but significant coupling to other
($n' \neq n$) vibrational levels exists. When atoms are transferred from $V_{\da}\left( {\bf x}\right) $ to $
V_{\ua}\left( {\bf x}\right) $ by the microwave field, a parametric excitation of the atomic wavepacket is
induced, thereby exciting ``breathing'' modes. However, the change in the optical potential also changes the
gravitational sag, and hence excites ``sloshing'' modes along the vertical axis. A wavefunction parametrically
excited in an harmonic oscillator will revive after $\tau =m\times \half\,\tau _{osc}$ (for $m=1,2,\ldots$)
yielding $\left| \left\langle \varphi _{n}\left( t=0\right) \mid \varphi _{n}\left( t=m\times \half\,\tau
_{osc}\right) \right\rangle \right| =1$. A sloshing mode will revive after $\tau =m\times \tau _{osc}$
\cite{Buchkremer00,Ejnisman97,Raithel98}. These revivals are seen in the middle curve of figure
\ref{echo_revivals} as a partial revival in the echo signal at $\tau =1.8 \msec$ and a stronger one at $\tau =3.6
\msec$, exactly equal to the measured trap oscillation frequency in the radial direction. For the upper curve in
figure \ref{echo_revivals}, a faster initial decay of the echo coherence is seen, as expected for a stronger
perturbation, followed by a clear revival at $\tau \sim3.3 \msec$. The origin of these revivals is verified in
numerical calculations of the echo signal for typical single states in a 2D harmonic oscillator with gravity,
having the same oscillation frequency as our experimental data (see figure \ref{echo_sim}). A similar calculation
shows that without the effect of gravity, the revival at $\tau =1.8 \msec$ is complete. The lack of a perfect
revival at $\tau =\tau _{osc}$ in figure \ref{echo_revivals}, is due to the anharmonicity of the Gaussian trap.
Note, that our technique is sensitive enough to map the quantum dynamics of a system, due to a perturbation
approximately three orders of magnitude smaller than $k_BT$.

For a sufficiently long time $\tau $ the wave packet oscillations of figure \ref{echo_revivals} damp due to a
complete dephasing of the dynamics. At such a long time a simple expression for the echo signal can be given. In
particular, assuming random phases between all vibrational states yields the simple relation
\begin{eqnarray}
\left| \left\langle \varphi _{n}\left( t=0\right) \mid \varphi _{n}\left( t=\tau \right) \right\rangle \right|
=\left| \left\langle n'=n\mid n\right\rangle \right| ^{4}.
\end{eqnarray}
Substituting this into equation \ref{tcf} and averaging over the ensemble yields the expected long-time echo
signal. We performed this calculation numerically for a 2D harmonic trap, in gravity, with our measured
oscillation frequency and a thermal ensemble with a temperature of $20 \mck$ clipped at our trap depth of
1.5$k_{B}T$. The results are shown in figure \ref{echo_long_time}, together with the measured long-time echo
signals as a function of perturbation strength\footnote[1]{The ensemble average for the clipped thermal
distribution in the harmonic trap was calculated as: $\sum\limits_{n,m}\left( \left| \left\langle u_{m}^{v2}\mid
u_{m}^{v1}\right\rangle \right| \left| \left\langle u_{n}^{h2}\mid u_{n}^{h1}\right\rangle \right| \right) ^{4}
e^{-E_{nm}/k_{B}T} /\sum\limits_{n,m} e^{-E_{nm}/k_{B}T} $ where $u_{n}^{h1}$ are eigenfunction of the harmonic
oscillator associated with the horizontal motion and $\left| 1\right\rangle $%
, $u_{n}^{h2}$ are associated with $\left| 2\right\rangle $ (the spring constant is slightly stronger).
$u_{m}^{v2}$ and $u_{m}^{v1}$ are associated with the vertical motion where there is both a change in spring
constant and a change in zero point due to different gravitational sag. The sum is over the $\sim$3$\times10^{6}$
states for which $E_{nm}=\left( n+m\right) \hbar \omega_{osc}<U_{pot}$.}. As seen, the calculation for the
harmonic trap and the data points for the Gaussian trap show the same qualitative behavior, of improved long time
echo when $\varepsilon$ becomes small.

We end this section with a brief discussion of the validity of the ``no mixing'' condition
$\braket{n}{n'}=\delta_{n,n'}$ in our experimental system. We start by noting that typically $10^{10}-10^{12}$
states of our Gaussian optical trap are thermally populated (this is easily estimated by noting that we typically
trap $\sim 10^5$ atoms at the standard phase space density provided by laser cooling techniques, $\sim 10^{-6}$).
However, since the duration of the experiment is often much shorter than the longitudinal oscillation time of
atoms in our trap, this longitudinal motion is essentially frozen, and we can consider only the $10^6-10^8$
thermally populated transverse vibration states. Perturbation theory indicates that a necessary condition for
substantial mixing of states is that the energy shift of populated states is comparable to the energy spacing
between adjacent states. This means that for $10^6$ thermally populated states the relative energy shift needs to
be much smaller than $10^{-6}$ to ensure the ``no mixing'' condition. In our experiments we typically observe
mixing (i.e. a fast decay of the echo coherence) only for much larger relative perturbations of $\sim 10^{-3}$.
The solution for this ``paradox'' is that state mixing by a perturbation can be drastically suppressed by strong
selection rules, making the \emph{sufficient} condition for mixing $\sim 10^3$ times larger than the
\emph{necessary} one. For our Gaussian trap such selection rules are provided by the fact that the transverse
motion of the atoms is completely separable into two independent one-dimensional motions, with only $\sim 10^3$
one-dimensional states thermally populated in each direction. Orthogonality thus ensures mixing only between
motional states in each direction separately, whose nearest-state separation is $\sim 10^3$ times larger then for
the general two-dimensional states. Such $\times\,1000$ enhancement in the stability of the quantum states against
perturbation in our separable system is only one of many similarly dramatic difference of the quantum properties
between separable and non-separable dynamical systems. Further discussion of such differences exists in section
\ref{sec:chaos} and in references \cite{Andersen04b, Andersen04c}.

\section{``Multiple $\pi$ Pulses'' Sequence} \label{ssec:manypi}
We turn now to investigate the limitations on the coherence time achieved by the echo scheme. According to the
time-independent Hamiltonian of equation \ref{ham} the echo scheme is expected to provide a complete cancellation
of the dephasing for arbitrarily long times, provided the condition $ \left\langle n^{\prime }\mid n\right\rangle
\simeq \delta _{n,n^{\prime }}$ is fulfilled for all thermally populated states. However, in real physical systems
this is not the case \cite{Kuhr03}. Two types of mechanisms cause the echo coherence to decay for long times even
when $ \left\langle n^{\prime }\mid n\right\rangle \simeq \delta _{n,n^{\prime }}$. The first type of mechanisms
(which we denote as ``dynamical'' $T_{2}$ processes) are time-dependent processes leading to a time-dependent
resonance frequency of the two-level system. This will cause the two parts of the wave function generated by the
first $\pit$-pulse to acquire different phases during the two ``dark'' periods of time $\tau $, hence causing
imperfect interference at the time of the second $\pit$-pulse. Examples of such mechanisms are fluctuations of the
trap depth e.g. due to noise in the trap laser power \cite{Kuhr03}, fluctuations in the bias magnetic field giving
rise to a fluctuating second order Zeeman shift, and spontaneous Rayleigh scattering of a photon from the trap
laser. Rayleigh scattering of photons does not lead to instantaneous loss of coherence, as does a Raman scattering
event \cite{Cline94}. Nevertheless, the recoil energy acquired by the atom can significantly change its
vibrational level, and therefore its resonance frequency. The result is, again, a time-dependent resonance
frequency of the two level systems. Other heating mechanisms, such as pointing instability of the trap laser beam,
typically involve a much smaller energy change than Rayleigh photon scattering, hence they induce a much longer
time scale for dephasing. The second type of mechanisms for decay of echo coherence ($ T_{1}$ processes) relates
to the finite lifetime of the internal states of the atoms, which is limited mainly by transitions induced by the
trap laser light.

\begin{figure}[tbp]
\begin{center}
\includegraphics[width=3in]{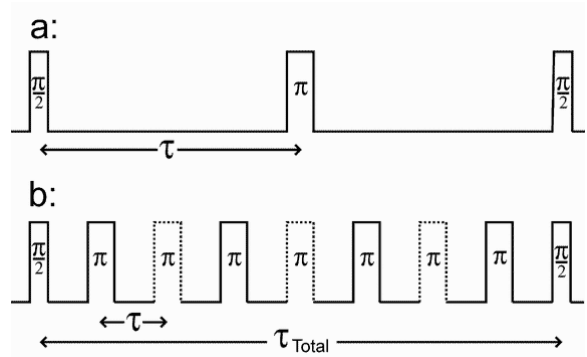}
\end{center}
\caption{a: Echo pulse sequence. b: ``Multiple-$\pi$'' pulse sequence. Shown in dashed lines are $\pi$ pulse which
occur at times where phases are compensated, hence they are omitted, with the advantage of less coupling to
$m_F\neq 0$ states, as explained in the text. In our experiment an additional $\pi$-pulse (not shown in the
figure) is used immediately following the $\pit$-pulse at the end of the sequence, in order to measure decoherence
as a deviation from $P_{\ua}$=0.} \label{sequences}
\end{figure}

As stated above, dynamical $T_{2}$ processes do not cause instantaneous decoherence, but are characterized by a
typical time scale over which a substantial phase difference evolves. If the time between pulses $\tau $ is larger
than this time scale, we expect the coherent signal to disappear. The time $\tau $ can be reduced by adding more
(equally spaced in time) $\pi $-pulses between the two $\pit$ -pulses (see figure \ref{sequences}). If the decay
of the echo coherence is dominated by dynamical $ T_{2} $ processes, and the time scale for variations in the
resonance frequency is longer than $\tau$, we expect a coherent signal to reappear \cite{Andersen04}. The method
has a strong similarity with the ``decoupling'' method used in quantum information schemes in NMR, where a
repeated fast refocusing completely stops the dynamics (See reference \cite{Gershenfeld97} and references
therein).

In this experiment the dipole trap consists of a $370 \mw$ linearly polarized horizontal Gaussian laser beam
focused to a 1/e$^{2}$ radius of $50 \mcm$ and with a wavelength of $\lambda=810 \nm$, yielding a trap depth of $
U_{0}\approx 100 \mck$. The transverse oscillation time of atoms in the trap is measured by parametric excitation
spectroscopy to be $1.4 \msec$ \cite{Friebel98} ensuring $ \left\langle n^{\prime }\mid n\right\rangle \simeq
\delta _{n,n^{\prime }}$, for all thermally populated transverse vibrational states. The bias magnetic field is
stronger ($250 \mg$) and turned on after the atoms are loaded into the trap.

\begin{figure}[t]
\begin{center}
\includegraphics[width=2.87in]{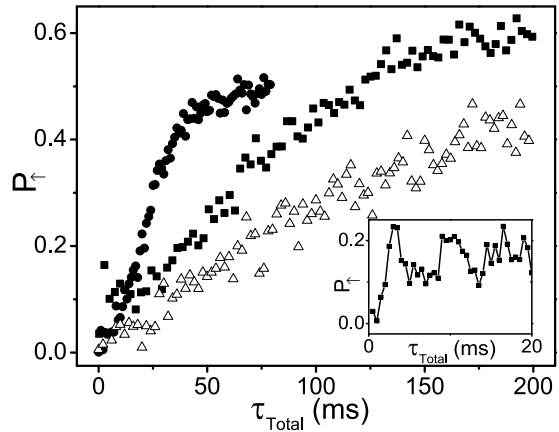}
\end{center}
\caption{$\bullet$ : Coherence signal ($P_{\ua}$) for the echo pulse sequence (figure \ref{sequences}a) as a
function of $\tau _{Total}$. $\blacksquare$ : $P_{\ua}$ for a pulse sequence with six $\pi$-pulses (figure
\ref{sequences}b) as a function of $\tau _{Total}$. It is seen that adding more $\pi$-pulses increases the
coherence time, but also leads to an initial small and rapid partial coherence decay. In the inset the short time
signal for a pulse sequence with ten $\pi$-pulses is shown. Wave-packet revivals are seen for $\tau_{Total}=7$ and
$14 \msec$. $\vartriangle$ : $P_{\ua}$ for a $\pi$-$\pi$ pulse sequence as a function of $\tau _{Total}$. A
monotonic increase in the signal is seen, due to transitions between m$_{F}$-states within the same $F$-manifold.
} \label{echo_multi}
\end{figure}

As seen in figure \ref{echo_multi}, the echo signal starts from $P_{\ua}=0$ (indicating perfect coherence) and
monotonically approaches $P_{\ua}=\frac{1}{2}$ (indicating complete loss of coherence). The ($1/e$) coherence time
$\tau _{c}$ is seen to be $\tau _{c}=26 \msec$. This coherence time is already $\sim 5$ times longer than the
Ramsey fringes decay time, showing the strong suppression of dephasing provided by the echo technique
\footnote{The trap in the experiment reported here is $\sim$ 4.5 times deeper than the trap used in section
\ref{sec:echo}, hence the echo coherence time is shorter. The Ramsey decoherence time, which is determined mainly
by the temperature of the atoms, is almost equal.}. Next, we add more $\pi $-pulses using the pulse sequence shown
in figure \ref{sequences}. First a $\pit$-pulse creates a coherent superposition state of $\ket{\da}$ and
$\ket{\ua}$. After time $\tau $ the first $\pi $-pulse is applied, followed by the rest of the $\pi $-pulses at
time intervals $2\tau $. A second $\pit$-pulse is applied at time $\tau$ after the last $\pi $-pulse, and followed
immediately by an additional $\pi $-pulse (not shown in the figure) in order to have $P_{\ua}=0$ for a coherent
signal for the even number of $ \pi $-pulses that we use. The signal as a function of the total time between the
first and the last pulse, $\tau _{Total}$, for a pulse sequence containing six $\pi $-pulses is also shown in
figure \ref{echo_multi}. As seen in the graph, at long times $P_{\ua}$ exceeds 1/2, which corresponds to total
decoherence for a purely 2-level system. We attribute this to a weak population mixing between $m_{F}$-states
within the same $F$-manifold as a consequence of spontaneous Raman scattering and off-resonant stimulated Raman
transitions involving two photons from the trap laser. The net effect of these transitions is to increase the
population of the $F=3$ manifold after the subsequent microwave pulses. To isolate and directly measure this
population mixing effect, we measure the population of $F=3$ as a function of time between two
population-inverting $\pi $-pulses \footnote{To further support our interpretation, we verify that a reduction in
the transition rate for larger values of the bias magnetic field is observed. Moreover, we verify that the
asymptotic slope of the multiple-$\pi$ echo signal is indeed half the slope of the $\pi$-$\pi$ measurement, as
expected from the fact that the $m_{F}$ population difference of the former is roughly half that of the latter.}.
From the results, also presented in figure \ref{echo_multi}, we measure the $F$-conserving transition rate to be
$1.2 \seconds^{-1}$. Combining these results with the (independently measured) $0.6 \seconds^{-1}$ rate of
$F$-changing transitions \cite{Cline94,Ozeri99}, yields a population decay rate of $1.8 \seconds^{-1}$ for the
$m_{F}$=0 states.

Since the value of $P_{\ua}$ at long times is greater than 1/2 we conservatively define the coherence time as the
time $\tau _{Total}$ between the two $\pit$-pulses where $P_{\ua}$ reaches a value of $P_{\ua}=\frac{1
}{2}(1-\frac{1}{e})$. It is seen that $\tau _{c}=65 \msec$, clearly showing that the additional $\pi $-pulses
substantially increase the coherence time.

\begin{figure}[tbp]
\begin{center}
\includegraphics[width=3.5in]{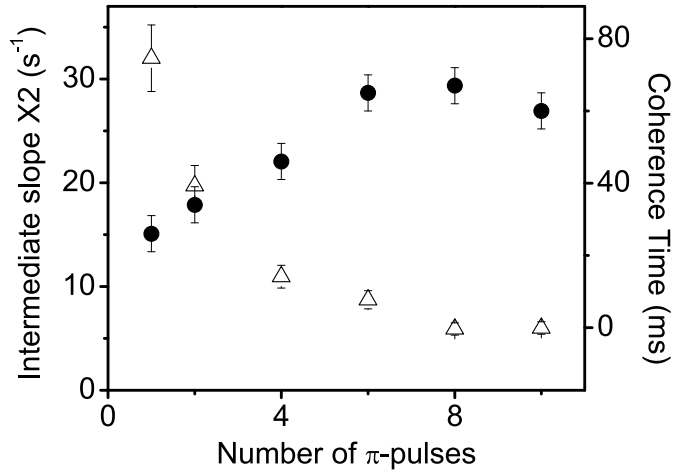}
\end{center}
\caption{($\bullet$) Coherence time as a function of the number of $\pi$-pulses. A maximum coherence time of
$\sim65$ ms is observed, as a consequence of a trade-off between suppressing $T_{2}$-processes and the increased
mixing to different vibrational states. ($\vartriangle$) Twice the slope of the intermediate regime, corrected for
contributions from $F$-conserving transitions.} \label{multi_number}
\end{figure}

As shown in figure \ref{echo_multi}, the multiple-$\pi$ sequence suffers from an initial small and rapid partial
loss of coherence. This is because $\left\langle n^{\prime }\mid n\right\rangle $ is not strictly a delta
function, hence when more $\pi $-pulses are added, the mixing to other vibrational levels is increased and the
dynamic effects discussed in section \ref{ssec:qdynamic} appear (The asymptotic value of $P_{\ua}$ due to this
mixing effect is $\frac{1}{2}[1-\left\langle n^{\prime }\mid n\right\rangle^{2(n_{\pi}+1)}] $ where $n_{\pi}$ is
the number of $\pi $-pulses. The effect is too weak to be observed for $n_{\pi}=1$). Shown in the inset of figure
\ref{echo_multi}, is the short time coherence signal for a sequence with 10 $\pi$-pulses. Wave packet revivals
appear for $\tau _{Total}=$7 and 14 ms (i.e. a time between $\pi $-pulses of 0.7 and 1.4 ms) as expected for an
harmonic trap with our measured transverse oscillation time of 1.4 ms.

In figure \ref{multi_number} the coherence time $\tau _{c}$ is shown as a function of the number of $\pi $-pulses.
It is seen that the coherence time initially shows linear dependence on the number of $\pi$-pulses and then
reaches a maximum value of $\sim65 \msec$, a 2.5 fold improvement as compared to the simple echo coherence time.
The maximum coherence time is given by a trade-off between suppressing the dephasing due to $T_{2}$ processes, by
adding more $\pi$-pulses, and the increased dephasing due to mixing to other vibrational states, that the
additional $\pi$-pulses induce.

The decay rate after the initial fast decay shows the residual decoherence, not suppressed by the multiple-$\pi$
scheme. We fit the slope of the intermediate, nearly linear section of the coherence decay curves with a straight
line, and correct for the long time slope (We show in figure \ref{multi_number} twice the slope since the decay of
coherence contributes only 1/2 to $P_{\ua}$). We see that the addition of $\pi$-pulses improves the intermediate
slope by a factor of 6. However, as explained before, the initial mixing to other vibrational levels prevents us
from fully exploiting this improvement.

\section{Quantum Dynamics and Chaos}
\label{sec:chaos}

The coherence properties of trapped atoms are related to their dynamics in the trapping potential. Loosely
speaking, the fact that the same electron which is probed by the spectroscopy is used for the trapping, means that
there is an inherent interplay between dynamics and spectroscopy. Already at the first demonstration of hyperfine
spectroscopy in a dark optical trap \cite{Davidson95}, it was conjectured that dynamics and spectroscopy are
coupled in dipole traps, and that the classical trajectories of the atoms in the trap affect the coherence time of
a spectroscopic measurement performed on them. This  interplay between dynamics and spectroscopy means that
precision spectroscopy in optical traps can serve also as a sensitive probe of the dynamics.

The quantum manifestations of different types of classical dynamics is still an issue of unsettled debates, since
the unitary evolution characteristic of quantum mechanics, is inconsistent with the classical definition of chaos,
i.e. an exponential sensitivity to the initial conditions. A widely accepted signature of quantum chaos is the
correspondence between the spectral statistics of quantized classically chaotic systems and the one of the
canonical ensembles of random matrix theory \cite{Gutzwiller90}. An alternative approach is to look at the
stability of the dynamics with respect to small changes in the Hamiltonian, as first proposed by Peres
\cite{Peres84}. In this framework, the ``fidelity'' denotes the overlap between a state evolved by a Hamiltonian
$H_{\uparrow }$ with the same state evolved by a slightly perturbed Hamiltonian $H_{\downarrow }$ \cite{Peres84}.
Despite the renewed theoretical and numerical interest in the fidelity and its properties (see references
\cite{Cucchietti00,Jacquod01, Cerruti02,Cucchietti02b,Cucchietti02,Prosen02, Emerson02, Vanicek03,Wisniacki03,
Benenti03, Hiller04}), experimental studies in chaotic and mixed systems are still lacking, mostly due to the
difficulty of preparing highly-excited pure quantum states. The use of ``echo spectroscopy'' eliminates the need
for pure quantum state preparation, since quantum fidelities can be experimentally observed even in a thermal
ensemble and for extremely highly excited quantum states.

As a specific example, we use in reference \cite{Andersen03} ``echo spectroscopy'' to experimentally measure
quantum dynamics of ultra cold $^{85}$Rb atoms trapped in atom-optics billiards \cite{Milner01, Friedman01,
Kaplan01, Andersen02}, with underlying chaotic or mixed classical dynamics. In this experiment the trap is a
light-sheet wedge billiard, made from two crossed blue detuned light sheets defining the billiard walls and where
gravity confines the atoms in the vertical direction \cite{Davidson95}. The light sheets have in-focus ($1/e^2$)
dimensions of $20 \times 250$ $\mu m$, and by the use of cylindrical lenses mounted on rotational stages, the
wedge angle are adjusted in order to control the classical dynamics\footnote{The structure of phase space in an
ideal wedge billiard can be tuned from stability to chaos by varying the vertex half-angle, $\alpha$
\cite{Lehtihet86}. For $\alpha < 45^\circ$ phase space is mixed, and the size of the stable islands oscillates as
a function of $\alpha$, a behavior dubbed ``breathing chaos''. For $\alpha > 45^\circ$ the system is fully
chaotic.}. The confinement in the longitudinal direction is provided by the diffraction of the beams. The very
elongated shape of the trap allows us to consider only the transverse motion and neglect the longitudinal one,
which has a timescale much longer than the experiment time. The temperature of the atoms is much larger than the
mean level spacing in the trap, and the atoms typically occupy many (up to $\sim10^8$) states in the trap. The
measured echo signal is the ensemble average of all of them.

As shown in reference \cite{Andersen04b}, the echo amplitude can be expressed as a function of the local density
of states (LDOS). The LDOS denotes the local average of the absolute-value-squared matrix elements of the
transformation matrix from eigenstates of $H_{\downarrow }$ to eigenstates of $H_{\uparrow }$. Formally, it is
simply $\left| \left\langle m_{\uparrow }|n_{\downarrow }\right\rangle \right| ^{2}$ as a function of $m_{\uparrow
}-n_{\downarrow }$ averaged over a ensemble of neighboring $n_{\downarrow }$ with approximately the same energy.
The width over which the LDOS is nonvanishing is denoted the ``bandwidth'', and if it is large, we  expect a rapid
decay of the echo coherence. In the perturbative regime the LDOS displays system specific features, despite the
fact that the underlying classical dynamics is chaotic. In particular, it is evident from the semiclassical
calculations of \cite{Cohen01,Andersen04b} that for atom optics billiards, where the inherent perturbation is
localized on the billiard walls, the LDOS has pronounced peaks for $ m_{\uparrow }-n_{\downarrow }$ corresponding
to $E_{n}-E_{m}=h/\tau _{bl}$, where $\tau _{bl}$ is the typical time between encounters with the wall ($\simeq15$
ms in our experiments). This means that the echo signal will show partial revivals for $ \tau =\tau _{bl}$
\cite{Andersen04b}.

\begin{figure}[tb]
\begin{center}
\includegraphics[width=3.57in] {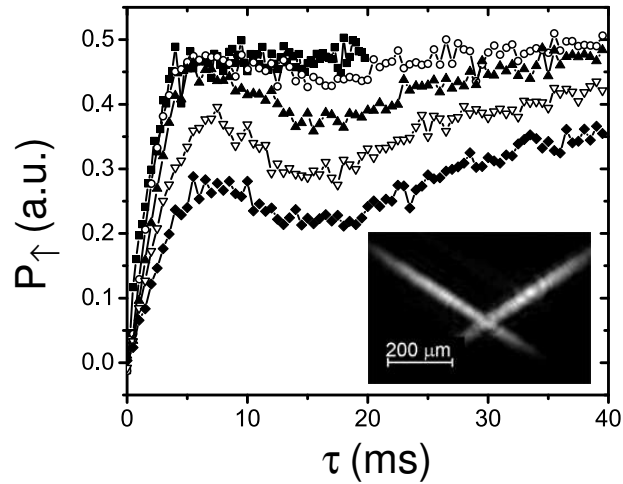}
\caption{Echo signal for a light sheet wedge with chaotic classical dynamics, for different perturbation strengths
(trap laser wavelength tuned from $\lambda=775.9$ to $\lambda=779.7$). $\blacklozenge $: $\epsilon=1.44 \times
10^{-3}$, $\triangledown $: $\epsilon=1.90 \times 10^{-3}$, $\blacktriangle $: $\epsilon=2.43 \times 10^{-3}$,
$\circ$: $\epsilon=3.80 \times 10^{-3}$ and $\blacksquare $: $\epsilon=1.52 \times 10^{-2}$. For small
perturbations a nonmonotonic decay with revivals around $\tau=\tau_{bl}$ is seen, whereas for large ones a
monotonic decay is observed. Inset: CCD image of the trap laser at the focal plane, showing a wedge with
$\alpha=52.5^\circ$).} \label{fig:echo_chaos}
\end{center}
\end{figure}

In figure \ref{fig:echo_chaos}, the decay of the echo signal for different perturbations is presented for a wedge
with vertex half-angle $\alpha=52.5^\circ$ (see inset) where the dynamics is fully chaotic. For small
perturbations a nonmonotonic decay is seen with a partial revival of correlations for $\tau \simeq \tau _{bl}$ as
predicted in \cite{Andersen04b}. The revivals are seen despite the fact that due to the ``high'' temperature  a
large band of energies is occupied in the trap. Partial revivals of correlations  in traps where the dynamics is
separable are generally expected at time scales equivalent to the 1D level spacing, but their surprising
observation here, for a trap with chaotic motion, demonstrates that they are a far more widespread phenomena.

For larger perturbations a crossover to a regime where the decay is monotonic is observed. In this regime the
decay is independent of perturbation strength. The perturbation is large enough so the overlap of equivalent
eigenstates is small, and this indicates that the effects of quantization of the trap levels should not play a
role and a classical description might be possible. Since the echo amplitude can be viewed as a propagator
\cite{Andersen04b}, and in our system the thermal de-Broglie wavelength is much smaller than the billiard's
dimensions, it is possible to use a semiclassical propagator \cite{Gutzwiller90}. It is then seen that the
classical trajectories contributing to the ensemble average of the echo amplitude, are those that after evolving
forward in time in $H_{\uparrow}$ and $H_{\downarrow}$, and then backwards in time in $H_{\uparrow}$ and
$H_{\downarrow}$, return to the vicinity of their initial position. These trajectories we divide into two types:
those that during this propagation hit the wall, and those that do not. Since $H_{\uparrow}$ and $H_{\downarrow}$
are highly different mainly in the vicinity of the wall, then the action integral along the first type of
trajectories will yield a very large phase and the contribution from these trajectories to the ensemble average of
the echo amplitude will average out. However since the second type of trajectories does not feel the difference
between the potentials it will retrace its forward propagation backwards in time causing the action integral to
vanish. These trajectories will give a perfect contribution. Therefore the echo amplitude in this regime simply
measures the classical probability that the particles have not yet hit the wall. Calculations of the echo signal
using this model yields good agreement with the experimental measurements.

\section{Summary}

In this tutorial we present our experiments on the microwave spectroscopy of the ground state hyperfine splitting
of optically trapped atoms. We introduce a number of schemes to overcome the inherent perturbation of the atomic
levels by the trapping potential.

First, we present a simple, classical model and use it to analyze the light-induced perturbations on optically
trapped atoms, and in particular the inhomogeneous broadening of the hyperfine splitting, for a number of trap
geometries. An ``optimal'' trap, achieved by minimizing the surface to volume ratio of the trap (to almost that of
a sphere) and by choosing the minimal wall thickness which is allowed by diffraction, is shown to allow an atomic
coherence time which is $\sim 200$ longer than existing single-beam dark traps, and $ \sim 1800$ times longer than
red-detuned traps.

Next, we demonstrate a scheme for eliminating the trap-induced inhomogeneous broadening of the transition, by
adding a weak ``compensating'' laser, spatially mode-matched with the trapping laser and with a proper detuning
and intensity. Despite being tuned close to resonance, this laser induces a negligible change in the dipole
potential, and does not considerably increase the spontaneous scattering rate. With the suppression of
inhomogeneous broadening, the atomic coherence time is now limited by the much smaller spontaneous scattering
time. Whereas in a conventional optical trap the ac Stark shift of the line center strongly depends on the
temperature of the atoms, which may drift considerably, in the compensated trap the suppression of the line shift
is equally effective for all temperatures. Hence, it provides a means of achieving a higher stability of the line
center than that achieved by simply stabilizing the trapping laser detuning and intensity. For relative
spectroscopic measurements, such as the proposed measurement of the electron's permanent electric dipole moment
(EDM) \cite{Chin01}, only stability (and not absolute accuracy) of the line center is of importance. For example,
for a 10 $\mu$K deep YAG-laser trap, and a compensating beam with a $15 \khz$ (time averaged) frequency stability,
locking the relative intensity between both beams to a $1:10^{-5}$ stability, will result in $\sim10^{-14}$
stability of the microwave line center.

Conversely, we present a method in which a long microwave pulse is used to select a narrow energy band from the
atomic ensemble, around any central energy. The rest of the atoms are expelled from the trap using an on-resonant
laser, without perturbing the selected atoms. With our method, the central energy can be chosen to maximize the
number of selected atoms by selecting the energy with the highest density of populated states.

We also demonstrate that a macroscopic coherence, lost as a consequence of dephasing of the different populated
vibrational levels, can be efficiently revived by stimulating an echo if the trap detuning is large enough so
$\left| \left\langle n^{\prime }=n\mid n\right\rangle \right| \simeq 1$ for all thermally populated states. This
suppression of dephasing due to perturbations induced by the trap yields a dramatic increase in the coherence time
for trapped atoms, that may find important applications for precision spectroscopy and quantum information
processing. We further demonstrate that echo spectroscopy can be used to experimentally map the quantum dynamics
of trapped atoms, showing a crossover between quantum and classical regimes. In this way it can serve as an
extremely sensitive experimental tool for investigating quantum chaos in atom-optics billiards even when more than
10$^{6}$ states are thermally populated, thereby avoiding the need of preparing highly excited pure quantum
states.

We demonstrate that the dephasing in microwave spectroscopy of optical trapped atoms due dynamical changes in the
trap parameters, or other similar processes, can be suppressed beyond the suppression offered by the simple echo,
by using an improved pulse sequence containing additional $\pi$-pulses. The achieved coherence time is limited by
increased mixing between transverse states and, to a lesser extent, the lifetime of the internal states of the
atoms. Both these factors are expected to be substantially smaller for a trap laser with much larger detuning,
such as in \cite{Kuhr03}. The demonstrated pulse sequence may also find use in precision spectroscopy of a
periodic effect, where the $\pi$-pulses can be synchronized with the period of that effect.

Finally, we show that ``echo spectroscopy'' can serve as an important tool for the study of the properties of the
fidelity (or Loschmidt Echo) and its relation to the underlying classical dynamics of a quantum system. The main
difficulty in \emph{experimental} studies of these matters, is the preparation of highly-excited pure quantum
states. This difficulty can be tackled by using echo spectroscopy, which eliminates the need for pure quantum
state preparation and allows the experimental observation of quantum fidelities even in a thermal ensemble and for
extremely highly excited quantum states.

A first work in this direction in our experimental system was done in references \cite{Andersen04b,Andersen04c}.
Using a perturbative treatment we show that the coherence of the echo scheme is a function of the survival
probability or fidelity of eigenstates of the atoms in the trap \cite{Andersen04b}. The echo coherence and the
survival probability display ``system specific'' features, even when the underlying classical dynamics is chaotic.
We performed echo spectroscopy in chaotic and mixed atom-optics billiards, as a function of the perturbation
strength, and observed two different regimes \cite{Andersen04c}. First, a perturbative regime in which the decay
of echo coherence is non-monotonic and partial revivals of coherence are observed, as opposed to the prediction
from random matrix theory. These revivals are more pronounced in traps with mixed dynamics as compared to traps
where the dynamics is fully chaotic. Next, for stronger perturbations, the decay becomes monotonic and independent
of the strength of the perturbation. In this regime no clear distinction can be made between chaotic traps and
traps with mixed dynamics.

Some of the ideas presented here, have been found useful by other experimental groups. A ``compensating''
technique has been used \cite{Haffner03} to compensate the Stark shift in an ion trap quantum processor. Recently,
two other groups working on quantum information processing with single ions and atoms have reported the use of
coherence echoes \cite{Rowe02, Kuhr03}.

The combined possibilities offered by our experimental system (control of dynamical phase-space and spectroscopy)
can have impact on other fields of study: First, the effect of interactions between the particles trapped in a
billiard is an interesting question in the context of semiconductor quantum dots. In atom-optic billiards, the
interactions can be tuned in a controlled way, by changing the density of the trapped atoms or by the use of a
magnetic field near a Feschbach resonance to change the atom's scattering length \cite{Vogels97}. Next, open
systems, which are characterized by a scattering matrix (and not by eigenstates) can be investigated, and related
to results in quantum dots, such as universal conductance fluctuations \cite{Baranger98}. Finally, chaotic
atom-optic billiards, which their unique ability to move from the quantum to the classical regime, can be used for
experimental studies of parametrically-dependent Hamiltonians \cite{Cohen01b} and driven quantum systems
\cite{Cohen00b}.

The work described in this tutorial was supported in part by the Israel Science Foundation and the Minerva
Foundation.


\end{document}